\documentclass[american,aps,pra,amsmath,amssymb,showpacs, twocolumn]{revtex4-1}
\usepackage{colortbl}
\usepackage[T1]{fontenc}
\usepackage[utf8]{luainputenc}
\usepackage{amsmath}
\usepackage{amssymb}
\usepackage{xcolor}
\usepackage{graphicx}
\usepackage{verbatim}
\usepackage{multirow}
\usepackage{setspace}

\pdfpageattr {/Group << /S /Transparency /I true /CS /DeviceRGB>>}

\usepackage{mathbbol}
\usepackage{bbm}
\usepackage{dsfont}

\def\ket#1{|#1\rangle}
\def\bra#1{\langle#1|}
\def\ketbra#1#2{|#1\rangle\langle#2|}

\def\psiS{\Psi_S}

\DeclareMathOperator{\tr}{tr}

\begin{document}

\title{Anticoherence of spin states with point group symmetries}
\author{D.~Baguette$^1$, F.~Damanet$^1$, O.~Giraud$^2$, and J.~Martin$^1$}
\affiliation{$^1$Institut de Physique Nucl\'eaire, Atomique et de
Spectroscopie, Universit\'e de Li\`ege, B\^at.\ B15, B - 4000
Li\`ege, Belgium}
\affiliation{$^2$LPTMS, CNRS, Univ.~Paris-Sud, Université Paris-Saclay, 91405 Orsay, France}
\date{\today}
\begin{abstract}
We investigate multiqubit permutation-symmetric states with maximal entropy of entanglement. 
Such states can be viewed as particular spin states, namely anticoherent spin states. Using the Majorana representation of spin states in terms of points
on the unit sphere, we analyze the consequences of a point-group symmetry in their arrangement on the quantum properties of the corresponding state. We focus on the identification of anticoherent states (for which all reduced density matrices in the symmetric subspace are maximally mixed) associated with point-group symmetric sets of points. We provide three different characterizations of anticoherence, and establish a link between point symmetries, anticoherence and classes of states equivalent through stochastic local operations with classical communication (SLOCC). We then investigate in detail the case of small numbers of qubits, and construct infinite families of anticoherent states with point-group symmetry of their Majorana points, showing that anticoherent states do exist to arbitrary order.
\end{abstract}

%03.65.Aa	Quantum systems with finite Hilbert space
%Entanglement and quantum nonlocality, 03.65.Ud
	%in quantum information 03.67.Mn
% Quantum entanglement, 03.65.Ud

\maketitle

\section{Introduction}

Geometrical representations in science have a long history. They give further insight in many different contexts, ranging from classical mechanics to graph theory and quantum information~\cite{Zyczkowski_book}. One such representation is the Bloch sphere picture for spin-1/2 states, which has been widely used in the latter context. Several geometrical generalizations of the Bloch representation to higher spin systems have been proposed~\cite{Majorana,Mes10,Goy11,Man14,Gir15}. In his seminal paper~\cite{Majorana}, Majorana introduced a particularly convenient way of visualizing pure spin-$j$ states as a set of $2j$ points on the Bloch sphere. This representation has been used in various contexts, such as in the study of spinor Bose gases~\cite{Bar06,Sta13,Mak07} or entanglement quantification and classification in multiqubit systems~\cite{Martin10, Aulbach10, Mar11, Rib11}. It can be used to highlight symmetries of spinor wave functions and to determine the inert states of spin systems, which are stationary for generic energy functionals~\cite{Fiz12,Mak07}, and also to study the Berry phase acquired during a cyclic evolution of a spin system~\cite{Bru12}.

Spin-$j$ coherent states $\ket{\mathbf{n}}$ take a very simple form in the Majorana representation. These states are defined as the pure states verifying $\langle \mathbf{n} |{\mathbf{J}}|\mathbf{n}\rangle = j \mathbf{n}$ for some unit vector $\mathbf{n}$ ($\hbar$ is set to $1$), where ${\mathbf{J}}=({J}_x,{J}_y,{J}_z)$ is the spin angular momentum operator associated to a spin-$j$ system. In the Majorana representation, they are depicted by a single point ($2j$-fold degenerated) on the Bloch sphere, and are in this sense highly directional. For this reason, spin-coherent states are also called polarized states. By contrast, pure spin-$j$ states $\ket{\psi_j}$ that depart the most from coherent states could be defined as the states which do not display any directional properties. However, for any finite $j$, the moments $\langle({\mathbf{J}}\boldsymbol{\cdot}\mathbf{n})^k\rangle\equiv \bra{\psi_j}({\mathbf{J}}\boldsymbol{\cdot}\mathbf{n})^k\ket{\psi_j}$ of the spin components, from a certain order, depend on $\mathbf{n}$. At best, one can ask for the absence of directionality in the moments up to a given order. This motivates the following definition~\cite{Zimba_EJTP_3}: a spin-$j$ state $\ket{\psi_j}$ is said to be anticoherent to order $t$ (or $t$-anticoherent) if $\langle({\mathbf{J}}\boldsymbol{\cdot}\mathbf{n})^k\rangle$ is independent of the unit vector $\mathbf{n}$ for $k = 1, \ldots, t$. For example, $1$-anticoherence means that the expectation value of the spin operator ${\mathbf{J}}$ vanishes.

The problem of identifying anticoherent states can be tackled from a quantum information perspective. Quantum information mainly deals with the manipulation of information stored on a set of two-level quantum systems, or qubits. It is possible to rephrase the above definition of anticoherence in the qubit language. The key element is the one-to-one mapping between the Hilbert space of a single spin-$j$ (of dimension $2j+1$) and the symmetric subspace of the Hilbert space associated to a system of $N$ qubits (of dimension $N+1$) when $N=2j$. In this picture, spin-coherent states are fully separable, whereas symmetric $N$-qubit $t$-anticoherent states are characterized by the fact that their $t$-qubit reduced density matrices correspond to the maximally mixed state~\cite{DTJ14,Gir15}. As a consequence, $t$-anticoherent states are the most entangled symmetric states, in the sense that for any $(k, N-k)$ bipartition with $k = 1,\ldots, t$ the entanglement entropy is maximal. Note that the maximization of the entropy of entanglement over the \emph{whole} Hilbert space for all possible bipartitions is not reached for symmetric states~\cite{AME} and thus the states that realize this maximum are not $t$-anticoherent; however states maximizing the entanglement entropy for $(1, N-1)$ bipartitions can be symmetric and thus $1$-anticoherent. These states, with maximally mixed one-qubit reductions, have been the subject of numerous research~\cite{Gisin_PLA_246,Scott04,Gour_13,Brown_JPA_28,Arn13,Goy14,Mac15,Saw13}, notably due to their importance in two-party communication tasks. Similarly, the notion of non-classicality of spin states based on the Glauber-Sudarshan $P$ representation~\cite{Gir08,Lui11} has been translated to multiqubit symmetric states in~\cite{Dev13} where it has been shown to imply entanglement. 

The aim of this paper is to investigate anticoherence from a geometrical point of view, using the Majorana representation. As previously mentioned, for coherent states, Majorana points all lie at the same place. By contrast, states whose Majorana points are spread out as far away as possible on the sphere have been shown to be highly entangled~\cite{Martin10,Aulbach10,Cran10}. For instance, it was found that states obtained by considering Majorana points with a configuration identical to that of electric charges at equilibrium on a sphere maximize the geometric measure of entanglement for certain spin values. Such geometric arrangements are known to display point group symmetries. We can therefore expect that arrangements of Majorana points with a certain point group symmetry lead to highly entangled states. In this paper, we investigate along those lines the consequences of point group symmetries on anticoherence properties of spin states. As we will see, this allows us to construct states with high order of anticoherence.

The paper is organized as follows. In Sec~\ref{sec:def}, we present various characterizations of anticoherence, both from a spin and a multiqubit perspective. In particular, we derive a set of equalities that the components of a permutation-symmetric state must satisfy in order to be $t$-anticoherent. In Sec~\ref{sec:pointgroup}, we establish the consequences of the symmetries of Majorana points on the components of the corresponding state and determine wich symmetries lead to anticoherence. In Sec.~\ref{sec:slocc}, we show that any state displaying cyclic symmetry can be transformed under stochastic local operations and classical communications (SLOCC) to an anticoherent state, and establish a link between symmetries, anticoherence and SLOCC classes. We also identify all anticoherent states with cyclic symmetry for $5$ qubits. In Sec.~\ref{sec:famillies}, we present a systematic method allowing us to find $t$-anticoherent states, and identify infinite families of anticoherent states with cyclic group symmetry. In Sec.~\ref{sec:allorder}, we prove that there exists anticoherent states to all orders.

\section{Definition and characterizations of anticoherence}\label{sec:def}

\subsection{Correspondance between spin-$j$ states and multiqubit symmetric states}
\label{spinqubit}

In this section we formalize the correspondence between single spin-$j$ states and multiqubit symmetric states and transpose the definition of anticoherence given in the introduction to multiqubit symmetric states. We first describe spin-$j$ states, then $N$-qubit symmetric states and finally their one-to-one mapping.

\subsubsection{Single spin-$j$ states}

Any pure spin-$j$ state $\ket{\psi_j}$ can be expanded in the standard angular momentum basis $\{\ket{j,m}:-j\leqslant m \leqslant j\}$ of joint eigenstates of $\mathbf{J}^2$ and $J_z$ as
\begin{equation}\label{standardexpansion}
\ket{\psi_j}=\sum_{m=-j}^j c_m\,\ket{j,m}
\end{equation} 
with $c_m$ complex coefficients such that $\sum_{m}|c_m|^2=1$. Coherent states are eigenstates of ${\mathbf{J}} \boldsymbol{\cdot} \mathbf{n}$ with eigenvalue $j$, where $\mathbf{n}=(\sin \theta \cos \varphi,\sin \theta \sin \varphi, \cos \theta)$ is the unit vector pointing along the direction specified by the angles $(\theta, \varphi)$ on the unit sphere.  For spin-$\frac12$, their general expression reads 
\begin{equation}\label{coh12}
\ket{\mathbf{n}}=\cos(\tfrac{\theta}{2}) \ket{\tfrac12,\tfrac12}+\sin(\tfrac{\theta}{2})e^{i\varphi} \ket{\tfrac12,-\tfrac12}.
\end{equation}
More generally, a spin-$j$ coherent state $\ket{\mathbf{n}}$ has expansion
\begin{equation}\label{eq:cohstate}
\ket{\mathbf{n}}=\!\sum_{m=-j}^j \!\sqrt{C_{2j}^{j-m}}[\cos(\tfrac{\theta}{2})]^{j+m} [\sin(\tfrac{\theta}{2})e^{i\varphi}]^{j-m} \ket{j,m}
\end{equation}
where $C_{2j}^{j-m}$ is a binomial coefficient.

The Husimi function of $\ket{\psi_j}$, defined by
\begin{equation}\label{eq:husimi_func}
Q(\theta,\varphi)
=|\langle\mathbf{n}|\psi_j\rangle|^2,
\end{equation}
is the probability to find the spin in the coherent state pointing along the direction $(\theta,\varphi)$~\cite{Aga81}.

\subsubsection{$N$-qubit symmetric states}
Any pure symmetric state $|\psi_S\rangle$ of $N$ qubits can be expressed as 
\begin{equation}
\label{Dickeexpansion}
|\psi_S\rangle = \sum_{k=0}^N d_k \ket{D_N^{(k)}}
\end{equation}
with $d_k$ complex coefficients such that $\sum_{k}|d_k|^2=1$ and $\{\ket{D_N^{(k)}}:0\leqslant k \leqslant N\}$ the (orthonormal) symmetric Dicke states defined in the computational basis as
\begin{equation}
\label{Dicke}
\ket{D_N^{(k)}} = \frac{1}{\sqrt{C_N^k}} \sum_{\sigma} \ket{\underbrace{0\ldots 0}_{N-k}\underbrace{1\ldots 1}_{k}},\nonumber
\end{equation}
where $k=0,\ldots,N$ is the number of $1$'s and the sum runs over all permutations of the qubits. In the particular case of a symmetric \emph{separable} state $\ket{\Phi_S}=\ket{\phi}^{\otimes N}$, where
\begin{equation}\label{singlequbit}
\ket{\phi}=\alpha\ket{0}+\beta\ket{1}
\end{equation}
is a single-qubit state with $\alpha,\beta\in\mathbb{C}$, the expansion (\ref{Dickeexpansion}) takes the form
\begin{equation}\label{eq:sepstate}
\ket{\Phi_S}=\sum_{k=0}^N \sqrt{C_N^k}\,\alpha^{N-k} \beta^k \ket{D_N^{(k)}}.
\end{equation}

The probability to find the pure symmetric state $|\psi_S\rangle$ in the separable state $\ket{\Phi_S}$ is given by
\begin{equation}\label{eq:sepoverlap}
Q(\alpha,\beta)=|\langle \Phi_S|\psi_S\rangle|^2.
\end{equation}

\subsubsection{One-to-one mapping}\label{ssec:onetoone}
A one-to-one correspondence between Eqs.~(\ref{standardexpansion})--(\ref{eq:husimi_func}) and (\ref{Dickeexpansion})--(\ref{eq:sepoverlap}) can be made by setting $k=j-m$, $N=2 j$, $\alpha=\cos(\tfrac{\theta}{2})$ and $\beta=\sin(\tfrac{\theta}{2})\,e^{i \varphi}$, and allows a formal equivalence between the standard basis and the Dicke basis
\begin{equation}\label{eq:equiv}
\begin{aligned}
\ket{D_N^{(k)}}& \;\leftrightarrow\; \ket{\tfrac{N}{2},\tfrac{N}{2}-k},\\[4pt]
\ket{j,m}& \;\leftrightarrow \;\ket{D_{2j}^{(j-m)}}.
\end{aligned}
\end{equation}
The spin-$j$ angular momentum operator ${\mathbf{J}}$ is formally equivalent to the collective spin operator ${\mathbf{S}}=({S}_x,{S}_y,{S}_z)$ associated with the $N$-qubit system. Here, we define ${S}_l=\frac12\sum_{i=1}^N {\sigma}_{l}^{(i)}$ ($l=x,y,z$) and ${\sigma}_{l}^{(i)}=\mathbb{1}\otimes\ldots\otimes\mathbb{1}\otimes \sigma_l\otimes\mathbb{1}\otimes\ldots\otimes \mathbb{1}$, where $\sigma_l$ are the Pauli operators ${\sigma}_{x}=\ket{0}\bra{1}+\ket{1}\bra{0}$, ${\sigma}_{y}=i(\ket{1}\bra{0}-\ket{0}\bra{1}$) and ${\sigma}_{z}=\ket{0}\bra{0}-\ket{1}\bra{1}$, and appear as the $i$th factor in the tensor product. More precisely, the matrix elements $\bra{j,m}{\mathbf{J}}\ket{j,m'}$ and $\bra{D_N^{(k)}}{\mathbf{S}}\ket{{D_N^{(k')}}}$ are equal for $k=j-m$, $k'=j-m'$ and $N=2j$. Spin-coherent states coincide with symmetric separable states, $\ket{\mathbf{n}} \,\leftrightarrow\, \ket{\Phi_S}$, in the sense that the spin-$j$ coherent state given by Eq.~(\ref{eq:cohstate}) can be seen as the $2j$-fold tensor product of the spin-$1/2$ coherent state~(\ref{coh12}).

\subsection{General conditions of anticoherence}\label{gencond}

A $t$-anticoherent pure spin-$j$ state $\ket{\psi_j}$ is defined by the fact that $\langle({\mathbf{J}}\boldsymbol{\cdot}\mathbf{n})^k\rangle$ is independent of the unit vector $\mathbf{n}$ for $k = 1, \ldots, t$. A criterion for anticoherence has been given by Bannai~\cite{Bannai_JPA_44} for any $j$ in terms of correlators of ${J}_x,{J}_y$ and ${J}_z$ and their powers. In particular, it was found as a corollary that a state is $1$-anticoherent iff $\langle{J_x}\rangle=\langle{J_y}\rangle=\langle{J_z}\rangle = 0$. It is $2$-anticoherent iff it is $1$-anticoherent and fulfills the additional conditions $\langle{J}_x^2\rangle=\langle{J}_y^2\rangle=\langle{J}_z^2\rangle$, and $\langle{J_k}{J_l}\rangle = 0$ for $k,l\in\{x,y,z\}$ and $k\neq l$, meaning that both the expectation value and the variance of the spin components are equal in all directions. For larger order of anticoherence $t$, these conditions quickly become cumbersome.

This definition of anticoherence can be naturally translated to the multiqubit setting. The mapping of the previous subsection leads to the following definition: a multiqubit symmetric state $|\psi_S\rangle $ is $t$-anticoherent if $\langle ({\mathbf{S}}\boldsymbol{\cdot}\mathbf{n})^k\rangle$ is independent of $\mathbf{n}$ for $k = 1, \ldots, t$. 

A simple characterization has been given in~\cite{DTJ14,Gir15}. Let $\rho_t$ be the $(t+1)\times(t+1)$ reduced density matrix of $\rho=|\psi_S\rangle\langle\psi_S|$ obtained by tracing over $N-t$ qubits and expressed in the Dicke basis $\{\ket{D_t^{(k)}}:k=0,\ldots,t\}$ spanning the $t$-qubit symmetric subspace. Note that since symmetric states are by definition invariant under permutation of the qubits, the $t$-qubit reduced states do not depend on the choice of the $t$ qubits. A state is $t$-anticoherent if and only if $\rho_t$ is proportional to the identity matrix, namely
\begin{equation}\label{rhotid}
\rho_t=\frac{\mathbb{1}_{t+1}}{t+1},
\end{equation}
where $\mathbb{1}_{t+1}$ is the $(t+1)\times (t+1)$ identity matrix. This implies that $t$-anticoherent states maximize the entanglement entropy $S=-\mathrm{tr}(\rho_t\log \rho_t)$ among symmetric states for any bipartition $(t,N-t)$ of the $N$ qubits.

In this section we derive three different characterizations of anticoherence to any order $t$. The first one is in terms of expectation values of convenient combinations of spin operators, the second is in terms of Dicke coefficients of a state and the third is based on the multipolar expansion of the Husimi function associated to a state.

\subsubsection{In terms of expectation value of spin operators}

Let $A(q)$ be the real numbers defined by
\begin{equation}\label{Aq}
A(q)=\frac{1}{N+1} \sum_{k=0}^{N} \left(\frac{N}{2}-k\right)^q
\end{equation}
for integers $q$. For $q=2$, we have $A(2)=\tfrac{1}{12}N(N+2)$ and for $q=4$, $A(4)=\tfrac{1}{20}(3N(N+2)-4)A(2)$.  Note that for all odd $q$, $A(q)=0$. We now show that a symmetric state $\ket{\psi_S}$ is $t$-anticoherent iff expectation values of the operators $S_{+}^r S_z^q $ with ${S}_{+} = {S}_{x}+ i {S}_{y}$ verify
\begin{equation}\label{diagat}
\bra{\psi_S} S_{+}^r S_z^q \ket{\psi_S}\equiv\langle S_{+}^r S_z^q \rangle= A(q) \,\delta_{r0}
\end{equation}
for $r=0,...,t$ and $q=0,...,t-r$, with $\delta_{r0}$ the Kronecker symbol. This set of equalities is much more compact than similar conditions found in~\cite{Bannai_JPA_44}. They are equivalent to the fact that the $t$-qubit reduced density matrix $\rho_{t}$ is maximally mixed for $t$-anticoherent states: Eq.~(\ref{diagat}) expresses equality of the diagonal entries of $\rho_{t}$ ($r=0$), and vanishing of all off-diagonal entries ($r\ne 0$).

To show this, we write each entry of the reduced density matrix $\rho_t$ as~\cite{DTJ14}
\begin{equation}\label{dellr}
(\rho_t)_{\ell, \ell+r}=\sum_{q=0}^{t-r}C_{\ell q}^{(r)} \,\langle {S}_{+}^r {S}_z^q \rangle,
\end{equation}
where $C^{(r)}$ for $r=0,\ldots,t$ are invertible square matrices of dimension $t+1-r$, which do not depend on the state $\ket{\psi_S}$. Equation (\ref{dellr}) follows from three facts : i) any operator which is a linear combination of $\ketbra{D_N^{(k+r)}}{D_N^{(k)}}$ for $k=0,\ldots,N-r$ can be expressed as a linear combination of the operators ${S}_{+}^r {S}_z^q$ with $q=0,\ldots,N-r$ (see Appendix B of~\cite{DTJ14}), ii)  any entry $(\rho_t)_{\ell,\ell+r}$ can be expressed as the expectation value $(\rho_t)_{\ell,\ell+r}=\langle\psi_S|\hat{O}_{\ell,\ell+r}|\psi_S\rangle$ of an operator $\hat{O}_{\ell,\ell+r}$ which is a linear combination of the operators $\ketbra{D_N^{(k+r)}}{D_N^{(k)}}$ for $k=0,\ldots,N-r$ (see Appendix C of~\cite{DTJ14}), and iii) the decomposition of $\hat{O}_{\ell,\ell+r}$ into operators ${S}_{+}^{r} {S}_z^{q}$, which according to point i) involves powers of spin operators such that $q\leqslant N-r$, in fact only involves terms with $q\leqslant t-r$.
Since $C^{(r)}$ is invertible, it follows from Eq.~(\ref{dellr}) that the vanishing of $\langle {S}_{+}^r {S}_z^q \rangle$ for all $q=0,\ldots t-r$ is equivalent to the vanishing of $(\rho_t)_{\ell, \ell+r}$. Thus Eq.~(\ref{diagat}) for $r\ne 0$ is equivalent to the fact that $\rho_t$ is diagonal. 

According to Eq.~(\ref{dellr}) for $r=0$, diagonal entries of $\rho_{t}$ are all equal to $1/(t+1)$ iff the vector of components $\langle {S}_{z}^q \rangle$ (with $q=0,\ldots,t$) is given by the constant vector $[C^{(0)}]^{-1}\,(1,\ldots,1)^T/(t+1)$. It remains to show that this vector is equal to the vector $(A(0),\ldots,A(t))$. We do this by evaluating $\langle \psi_S^A | ({\mathbf{S}}\boldsymbol{\cdot}\mathbf{n})^q |\psi_S^A \rangle$ for an arbitrary anticoherent state $ |\psi_S^A \rangle$. For any state $|\psi_S\rangle $, one has $\langle \psi_S|{\mathbf{S}}\boldsymbol{\cdot}\mathbf{n})^q|\psi_S\rangle =\tr[\rho U_{\mathbf{n}}{S}_{z}^q U^\dagger_{\mathbf{n}}]=\tr[U^\dagger_{\mathbf{n}} \rho U_{\mathbf{n}}{S}_{z}^q]$ where $U_{\mathbf{n}}=u_{\mathbf{n}}^{\otimes N}$ with $u_{\mathbf{n}}=\exp(-i\theta\,\mathbf{n}\boldsymbol{\cdot}\boldsymbol{\sigma}/2)$ the rotation operator of angle $\theta$ around the $\mathbf{n}$-axis. For a $t$-anticoherent state $ |\psi_S^A \rangle$ with density matrix $\rho^A$, the latter expression does not depend on the unit vector $\mathbf{n}$ for $q=0,\ldots,t$, and thus we can replace it by its average over all $\mathbf{n}$, so that
\begin{equation}\label{Snav}
\langle \psi_S^A | ({\mathbf{S}}\boldsymbol{\cdot}\mathbf{n})^q |\psi_S^A \rangle= \frac{1}{4\pi}\int \tr[U^\dagger_{\mathbf{n}'} \rho^A U_{\mathbf{n}'}{S}_{z}^q]\,d\mathbf{n}'.
\end{equation}
Using the completeness relation for coherent states, it is easy to show that for any density matrix $\rho$, we have $\bra{D_N^{(k)}}\int U^\dagger_{\mathbf{n}} \rho U_{\mathbf{n}} d\mathbf{n}\ket{D_N^{(k')}}=4\pi\,\delta_{kk'}/(N+1)$~\cite{Zyczkowski_book}.
Thus for any unit vector $\mathbf{n}$ and any $t$-anticoherent state $|\psi_S^{A}\rangle$ ($t\geqslant q$), we have
\begin{equation}
\langle \psi_S^A | ({\mathbf{S}}\boldsymbol{\cdot}\mathbf{n})^q |\psi_S^A \rangle=\frac{\tr[{S}_{z}^q]}{N+1}=A(q),
\end{equation} 
which completes the proof.

\subsubsection{In terms of Dicke coefficients}

The conditions for anticoherence of a state in terms of its Dicke coefficients $d_k$ (see Eq.~(\ref{Dickeexpansion})) are obtained by expressing Eqs.~(\ref{diagat}) in the Dicke basis. This gives the "diagonal" conditions ($r=0$)
\begin{equation}\label{diagatdk}
\sum_{k=0}^N \left(\frac{N}{2}-k\right)^q |d_k|^2 = A(q)
\end{equation}
for $q=0,\ldots,t$, and the "off-diagonal" conditions
\begin{equation}\label{nondiagatdk} 
\sum_{k=0}^{N-r} B(k,r) \left(\frac{N}{2}-k\right)^q d_k^* d_{k+r}= 0
\end{equation}
for $r=1,\ldots,t$ and $q=0,\ldots,t-r$, where
\begin{eqnarray*}
B(k,r) &=& \bra{D_N^{(k+r)}}{S}_{+}^r\ket{D_N^{(k)}} \\
&=& i^r\sqrt{(k+1)(k-N)(r-1)_{k+2}(r-1)_{k+N+1}}
\end{eqnarray*}
with $(x)_n=x(x+1)\ldots(x+n-1)$ the Pochhammer symbol.

For instance, Eqs.~(\ref{diagatdk})--(\ref{nondiagatdk}) yield, for $1$-anticoherence, the conditions
\begin{align}
& \sum_{k=0}^{N}(N-2k)\,|d_k|^2 = 0, \label{mngc1} \\
& \sum_{k=0}^{N-1}\sqrt{(N-k)(k+1)}\,d_k^*d_{k+1} = 0, \label{mngc2}
\end{align}
which were first derived in~\cite{DTJ14}. For $2$-anticoherence, in addition to Eqs.~(\ref{mngc1})--(\ref{mngc2}), the following conditions must be fulfilled:
\begin{align}
& \sum_{k=0}^{N}\left(\frac{N}{2}-k\right)^2|d_k|^2 = \frac{N}{12}(N+2),\label{t2c1}\\
& \sum_{k=0}^{N-1}\left(\frac{N}{2}-k\right)\sqrt{(N-k)(k+1)}\,d_k^*d_{k+1} = 0,\label{t2c2}\\
& \sum_{k=0}^{N-2}\sqrt{(N-k)(N-k-1)(k+1)(k+2)}\,d_k^*d_{k+2} = 0.\label{t2c3}
\end{align}
More generally, conditions of anticoherence to order $t$ are obtained by adding to those of order $t-1$ a set of $t+1$ equations, one given by (\ref{diagatdk}) with $q=t$ and the $t$ other equations given by (\ref{nondiagatdk}) with $q+r=t$. So the total number of (real) equations for $t$-anticoherence is $(t+1)^2-1$, as expected from condition (\ref{rhotid}).

\subsubsection{Based on $Q$ and $P$ functions}\label{gencondQP}

This characterization of anticoherence provides a more direct physical insight. The density matrix $\rho$ of a spin-$j$ state can always be expanded into state multipole operators $T_\ell^m$~\cite{Aga81} as
\begin{equation}\label{exprho}
\rho= \sum_{\ell=0}^{N} \sum_{m=-\ell}^{\ell} c_{\ell m} T_\ell^m.
\end{equation}
It can be further described in terms of its Husimi function $Q(\theta,\varphi)=\langle \mathbf{n} | \rho \ket{\mathbf{n}}$, or in terms of its Glauber-Sudarshan $P$ function
defined by
\begin{equation}
\rho=\int P(\theta,\varphi) \ket{\mathbf{n}}\bra{\mathbf{n}}d\mathbf{n}.
\end{equation}
Both $Q$ and $P$ functions can be expanded over spherical harmonics $Y_\ell^m(\theta,\varphi)$ with coefficients $Q_{\ell m}$ and $P_{\ell m}$ proportional to $c_{\ell m}$, see e.g.~\cite{Aga81}. As shown in \cite{Gir15}, a state is $t$-anticoherent iff $c_{\ell m}$ vanishes for all $\ell,m$ with $0<\ell\leqslant t$ and $-\ell\leqslant m\leqslant \ell$. From the proportionality of expansion coefficients of $Q$ and $P$ with $c_{\ell m}$, we deduce that for a $t$-anticoherent state all spherical multipole moments $Q_{\ell m}$ and $P_{\ell m}$ of order $\ell$ with $0<\ell\leqslant t$ of the Husimi and $P$ functions vanish.
As the Husimi function is the probability of finding the state in a coherent state with a specific direction, and as the spherical harmonics $Y_\ell^m(\theta,\varphi)$ are more and more oscillating as $\ell$ increases, directionality of a $t$-anticoherent state is less and less pronounced as $t$ increases because the Husimi function is more and more uniform on the sphere.

\section{Point groups for Majorana points}\label{sec:pointgroup}

\subsection{Majorana representation}
\label{Majorep}
A particularly convenient way of visualizing pure spin-$j$ states or $N$-qubit symmetric states is the Majorana or stellar representation~\cite{Majorana}. Any pure symmetric state $\ket{\psi_S}$ can be represented by a set of $N=2 j$ points with angles $\{(\theta_i,\varphi_i), i=1,\ldots,N\}$ on the Bloch sphere. These $N$ points are associated with $N$ pure single-qubit states $\ket{\phi_1},\ldots,\ket{\phi_N}$, where $\ket{\phi_i}=\cos{(\tfrac{\theta_i}{2})}\ket{0}+e^{i \varphi_i}\sin{(\tfrac{\theta_i}{2})}\ket{1}$, such that
\begin{equation}\label{overj}
\ket{\psi_S}=\mathcal{N}\sum_{\sigma}\ket{\phi_{\sigma(1)}\ldots \phi_{\sigma(N)}}
\end{equation}
where the sum runs over all permutations $\sigma\in S_N$ (the permutation group of $N$ elements) and $\mathcal{N}$ is a normalization constant. Let $n_S$ and $N-n_N$ be respectively the first and the last indices of non-vanishing Dicke coefficients $d_k$ for a state of the form (\ref{Dickeexpansion}). 
The Majorana representation \cite{Majorana} of such a state is obtained by finding the roots of the polynomial
\begin{equation}
\label{MajoranaPolynomialIdentity1}
P(z) = \sum_{k=n_S}^{N-n_N} (-1)^{k} \sqrt{C_N^k}\,d_k z^k .
\end{equation}
This polynomial can be put in the form
\begin{equation}
P(z)=(-1)^{N-n_N}\sqrt{C_N^{n_N}} d_{N-n_N}z^{n_S}\hspace{-.2cm}\prod_{k=1}^{N-n_N-n_S}\hspace{-.3cm}(z-z_k),
\label{MajoranaPolynomialIdentity2}
\end{equation}
where $z_k$ are the non-zero roots of $P(z)$. Applying the (inverse) stereographic projection from the complex plane onto the Bloch sphere to the zeroes of $P(z)$ through the relation $z_m=\cot(\theta_m/2)e^{-i\varphi_m}$ yields $N-n_S-n_N$ points $(\theta_m,\varphi_m)$, and $n_S$ points at the South pole of the Bloch sphere (corresponding to the roots located at $z=0$). Adding $n_N$ points at the North pole of the Bloch sphere yields the Majorana representation of $\ket{\psi_S}$ as $N$ points on the sphere, called Majorana points.  For instance, for a Dicke state $\ket{D_N^{(k)}}$, the Majorana representation is given by $N-k$ points at the North pole and $k$ points at the South pole of the Bloch sphere. For a spin-coherent (or symmetric separable $N$-qubit) state $\ket{\mathbf{n}}$ (see Eq.~(\ref{eq:cohstate})), the Majorana points are $N$ points located at $(\theta,\varphi)$. 

One of the advantages of the Majorana representation is its behavior under local unitary transformations (LU). Symmetric LU are transformations of the form $U^{\otimes N}$ where $U$ is a unitary operator acting on a single qubit. It corresponds, up to a global phase, to a rigid rotation of the Bloch sphere, and thus of the Majorana points, thereby preserving the symmetries of their arrangement.

\subsection{Majorana points for anticoherent states}

From its definition, anticoherence of a state $\ket{\psi_S}$ is preserved by symmetric LU. In particular, this means that it is only a feature of the relative arrangement of Majorana points. We expect Majorana points of an anticoherent state to be spread out over the sphere as evenly as possible. Indeed, anticoherent states should be as different as possible from any directional (polarized) state and thus have a Majorana representation as distinct as possible from $N$ points located at the same place. 

\begin{figure}
\begin{centering}
\includegraphics[width=5cm,clip=true]{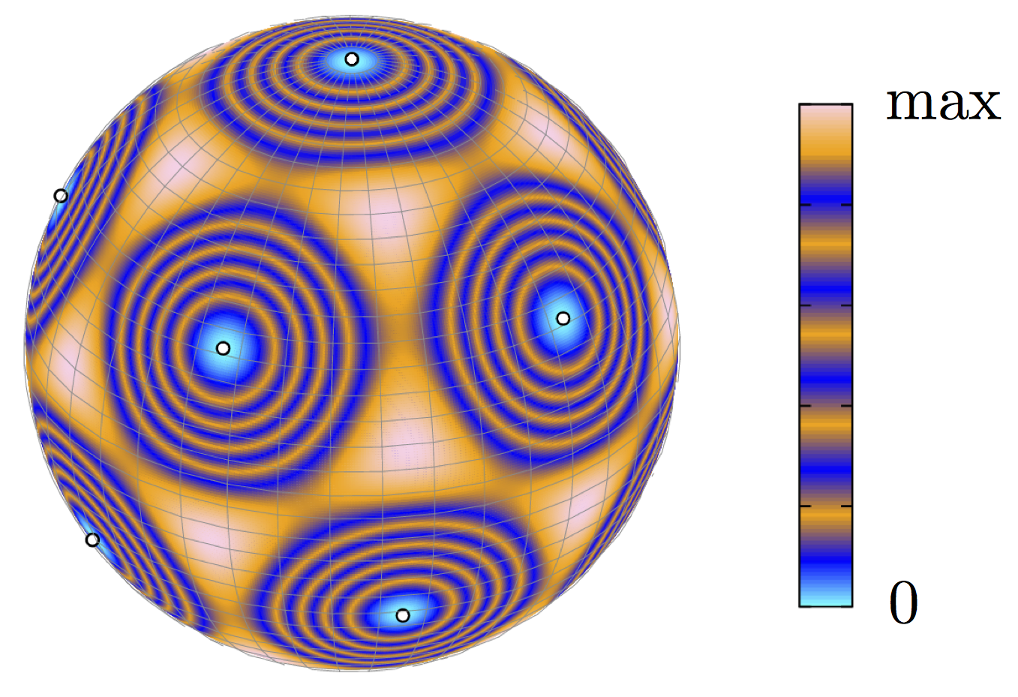}
\par\end{centering}
\caption{(Color online) Husimi function of the $12$-qubit $5$-anticoherent state $\ket{\psi_S}=\tfrac{\sqrt{7}}{5}\ket{D_{12}^{(1)}}+\tfrac{\sqrt{11}}{5}\ket{D_{12}^{(6)}}-\tfrac{\sqrt{7}}{5}\ket{D_{12}^{(11)}}$. The Majorana points, depicted by white points on the sphere, display an icosahedral symmetry. Color code is chosen so as to highlight the contour lines of the Husimi function.
 \label{Dn12} }
\end{figure}

As shown in sec.~\ref{gencondQP}, anticoherence is related to the vanishing of multipole moments of lower order and thus to the uniformity of the Husimi function over the sphere. A natural way of constructing states with vanishing multipole moments is to consider Majorana points arranged symmetrically on the sphere, as the zeroes of the Husimi function are diametrically opposite to the Majorana points on the Bloch sphere. For example, 12 points taken at the vertices of an icosahedron yield a 12-qubit state which is 5-anticoherent. Its Husimi function is depicted in Fig.~\ref{Dn12}. In the next subsection, we consider arrangements of points with symmetries belonging to the seven infinite point group families. Note that anticoherent states with symmetries corresponding to the exceptional point groups can also be constructed, as the example just mentioned shows.

As a consequence of these geometrical features, one may expect the barycenter of the Majorana points to coincide with the center of the Bloch sphere. In fact, this is not the case~\cite{DTJ14}. The distance from the barycenter to the center (which can serve to define an entanglement measure~\cite{Zyc12}) can be surprisingly large, as we show in the following example. States of the form $\ket{\psi_S}=\mathcal{N}(\sqrt{N-2}\ket{D_N^{((N-1)/2)}}+\ket{D_N^{(N-1)}})$ for odd $N>3$ have a Majorana representation corresponding to $(N-1)/2$ points at the South pole, $1$ point at the North pole and $(N-1)/2$ points arranged in a regular polygon parallel to the equatorial plane (a similar family can be found for even $N$). They are $1$-anticoherent, as can be checked from Eqs.~(\ref{mngc1})--(\ref{mngc2}); however, for large $N$, the barycenter of their Majorana points goes to $-1/5$. 

\subsection{Point-group symmetric states}\label{pgss}

Point groups are discrete subgroups of $O(3)$. There are seven infinite families of axial groups indexed by an integer $n$, and seven exceptional point groups.
The seven infinite point group families are generated by rotations and reflections. We can always bring by LU any point configuration with axial rotation symmetry to a configuration where the symmetry axis is the $z$-axis. We will denote by $r_{n}$ the rotation with axis $Oz$ and angle $2\pi/n$. Similarly, we can always bring a point configuration with reflection symmetry with respect to a plane containing the  $z$-axis to a configuration where this plane is the plane $Oxz$ containing both the $x$ and the $z$-axes. We denote by $\sigma_v$ the reflection with respect to this 'vertical' plane, and by $\sigma_h$ the reflection with respect to the 'horizontal' plane $Oxy$. Additional symmetry planes, traditionally denoted by $\sigma_d$, are obtained by rotating $\sigma_v$ by an angle $\pi/n$ around $Oz$.

The most fundamental family is that of cyclic groups $C_n$. The group $C_n$ is generated by the rotation $r_{n}$, i.e.~its elements are the $r_n^k$ with $0\leqslant k\leqslant n-1$. The other groups can be described as follows~(see e.g.~\cite{Cot90}) : 
$C_{nh}$, generated by $r_{n}$ and $\sigma_h$;
$C_{nv}$, generated by $r_{n}$ and $\sigma_v$ for odd $n$ and by $r_{n}$, $\sigma_v$ and $\sigma_d$ for even $n$;
$S_{2n}$, generated by the product $r_{2n}\sigma_h$;
$D_n$, dihedral group, generated by $r_{n}$ and by a rotation of angle $\pi$ around the $x$-axis which can be expressed as the product $\sigma_h\sigma_v$;
$D_{nh}$, generated by $D_n$ and $\sigma_h$;
$D_{nd}$, generated by $D_n$ and $S_{2n}$, or equivalently, by $r_{2n}\sigma_h$ and $\sigma_v$.

A given symmetry of the Majorana points of the state (\ref{Dickeexpansion}) reflects on its Dicke coefficients $d_k \in \mathbb{C}$. We first determine the consequences of a symmetry on the roots $z_k$ of the polynomial (\ref{MajoranaPolynomialIdentity2}). We recall that the mapping between the complex plane and the Bloch sphere is chosen as $z=\cot(\theta/2)e^{-i\varphi}$. If the Majorana representation of a state is invariant under a symmetry, then the product $z^{n_S}\prod_{k=1}^{N-n_N-n_S}(z-z_k)$ in (\ref{MajoranaPolynomialIdentity2}) must be left unchanged, and thus the polynomial (\ref{MajoranaPolynomialIdentity1}) remains unchanged up to a multiplicative constant. This induces relations between the Dicke coefficients. 

A rotation $r_n$ transforms $z$ into $z\exp(-2i\pi/n)$. Points at the poles are not affected by a rotation $r_n$. If a configuration of points is invariant under $r_n$, then 
\begin{eqnarray}
\prod_{k=1}^{N-n_N-n_S}\hspace{-.3cm}(z-z_k)&=&\prod_{k=1}^{N-n_N-n_S}\left(z-z_ke^{-2i\pi/n}\right)\\
&=&e^{-\frac{2i\pi}{n}(N-n_N-n_S)}\hspace{-.3cm}\prod_{k=1}^{N-n_N-n_S}\hspace{-.3cm}\left(ze^{\frac{2i\pi}{n}}-z_k\right).\nonumber
\end{eqnarray}
From Eq.~\eqref{MajoranaPolynomialIdentity2} we then get
\begin{equation}
\label{polycn}
P(z)=e^{-\frac{2i\pi}{n}(N-n_N)}P\left(ze^{\frac{2i\pi}{n}}\right).
\end{equation}
Using the expansion Eq.~\eqref{MajoranaPolynomialIdentity1} and identifying the coefficients in the polynomials on both sides of Eq.~\eqref{polycn} we get that for all $k$ with $n_S\leq k\leq N-n_N$
\begin{equation}
d_k=e^{-\frac{2i\pi}{n}(N-n_N-k)}d_k,
\end{equation}
unless $k=N-n_N+q n$.
In particular, for all $k$, we have that $d_k=0$ unless $k=n_N+q n$ with $q\in\mathbb{N}$. Because of the symmetry, $N-n_N-n_S$ has to be a multiple of $n$, so that the latter condition is equivalent to $k=n_S$ (mod $n$).

A similar approach can be followed for all other symmetry operations (see Appendix). The results are summarized in Table~\ref{tabDickeconstraints}.

\begin{table}[h!]
\renewcommand{\arraystretch}{1.6}
\begin{tabular}{|c|c|}
\hline 
\rule[-9.5pt]{0pt}{25pt} \parbox{1.6cm}{Symmetry\\operation} \rule[-9.5pt]{0pt}{25pt} & \hspace{6pt}Constraints on Dicke coefficients\hspace{6pt} \\
\hline 
\hline 
$r_n$ & $d_k=0\quad\forall \; k\ne n_S~(\mathrm{mod}~n)$ \\ 
\hline 
\multirow{2}*{$\sigma_h$}  &  $n_N=n_S$, and $\exists\, \xi\in\mathbb{R}:$\\
&  $d_{N-k}=e^{i\xi}d_k^* \quad\forall\,  k$ \\
\hline 
$\sigma_v$  &  $d_k\in\mathbb{R} \quad\forall \, k$\\ 
\hline 
$\sigma_d$  &  $d_k=\pm |d_k|e^{\frac{ik\pi}{2n}}\quad\forall \, k$ \\ 
\hline 
\multirow{2}*{$\sigma_h\sigma_v$}  &  $d_{N-k}= d_k\quad\forall \, k$, or \\ 
& $d_{N-k}=-d_k\quad\forall \, k$ \\ 
\hline 
\multirow{3}*{$s_n$}  & $n_N=n_S$, and $\exists\, \xi\in\mathbb{R}:$\\
& \rule[-24pt]{0pt}{25pt}\parbox[t]{5cm}{$d_k=0 \quad\forall  \, k\ne n_S~(\mathrm{mod}~n)$\\[6pt]
$d_{N-k}=(-1)^q e^{i\xi}d_k^* \quad$otherwise}\rule[-24pt]{0pt}{25pt}\\ 
\hline 
\end{tabular}
\caption{Constraints on the Dicke coefficients of a state $\ket{\psi_S}$ invariant under a certain symmetry operation on its Majorana points.}
\label{tabDickeconstraints}
\end{table}

%--------------------------------------------------------------------------
\subsection{Canonical form of point-group symmetric states}
\label{Canonical}
%--------------------------------------------------------------------------

From the results of Table~\ref{tabDickeconstraints}, we obtain a canonical form for the Dicke coefficients of a state whose Majorana points display a certain symmetry. That is, all states with point-group symmetric Majorana point arrangements can be brought to one of the forms (\ref{DickeCn})--(\ref{DickeDndm}), depending on the symmetry.

\subsubsection{Cyclic groups}

A state with $C_n$ symmetry is characterized by an arrangement of Majorana points invariant under $r_n$. From Table~\ref{tabDickeconstraints}, we obtain for the vector of Dicke coefficients $(d_0,d_1,\ldots,d_N)$ the canonical form
\begin{equation}
\label{DickeCn}
\mathbf{d}_{C_n} = (\mathbf{0}_{n_S},d_{n_S},\mathbf{0}_{n-1},d_{n_S+n},\mathbf{0}_{n-1},d_{n_S+2n},\ldots,\mathbf{0}_{n_N}),
\end{equation}
where $\mathbf{0}_m$ stands for a string of $m$ zeroes. The consequences of this symmetry are particularly interesting. Indeed, due to the particular form of the Dicke coefficients in Eq.~(\ref{DickeCn}), the $t$-qubit reduced density matrix is a diagonal matrix for $t<n$ since all off-diagonal conditions~(\ref{nondiagatdk}) are satisfied. As all the symmetry groups described in sec.~\ref{pgss} have $C_n$ as a subgroup, $t$-anticoherence with $t<n$ for point-group-symmetric states only requires the additional diagonal conditions~(\ref{diagatdk}) for $q=0,\ldots,t$.
 
For a state with $C_{nh}$ symmetry, the arrangement of points is additionally invariant under $\sigma_h$. The form (\ref{DickeCn}) together with the condition of invariance under $\sigma_h$ in Table~\ref{tabDickeconstraints} leads to the canonical form
\begin{multline}
\label{DickeCnh}
\mathbf{d}_{C_{nh}} = (\mathbf{0}_{n_S},d_{n_S},\mathbf{0}_{n-1},d_{n_S+n},\mathbf{0}_{n-1},d_{n_S+2n},\ldots\\
\ldots,\mathbf{0}_{n-1},d_{n_S+n}^*e^{i\xi},\mathbf{0}_{n-1},d_{n_S}^*e^{i\xi},\mathbf{0}_{n_S})
\end{multline}
with $\xi\in\mathbb{R}$.
For $n \geqslant 2$, states of the form (\ref{DickeCnh}) satisfy Eqs.~(\ref{mngc1})--(\ref{mngc2}) and are thus all $1$-anticoherent.

For a state with $C_{nv}$ symmetry, the conditions of invariance under $\sigma_v$ given in Table~\ref{tabDickeconstraints} and the form (\ref{DickeCn}) lead to the canonical form
\begin{multline}
\mathbf{d}_{C_{nv}} = (\mathbf{0}_{n_S},d_{n_S},\mathbf{0}_{n-1},d_{n_S+n},\\\mathbf{0}_{n-1},d_{n_S+2n},\ldots,\mathbf{0}_{n_N}),\, d_k\in \mathbb{R}.
\end{multline}

\subsubsection{Rotation-reflection group}
As $C_n$ is a subgroup of $S_{2n}$, any state with Majorana points displaying $S_{2n}$ symmetry is of the form (\ref{DickeCn}). The more general condition is that of invariance under $s_n$, which can be expressed as $n_S=n_N$, $N=2n_S+m n$ with $m$ even (for odd $m$, there is no way to achieve $S_{2n}$ symmetry), see~Table~\ref{tabDickeconstraints}. The canonical form is thus
\begin{multline}
\label{DickeS2n}
\mathbf{d}_{S_{2n}} = (\mathbf{0}_{n_S},d_{n_S},\mathbf{0}_{n-1},d_{n_S+n},\mathbf{0}_{n-1},d_{n_S+2n},\ldots\\
\ldots,\mathbf{0}_{n-1},-d_{n_S+n}^*e^{i\xi},\mathbf{0}_{n-1},d_{n_S}^*e^{i\xi},\mathbf{0}_{n_S})
\end{multline}
with $\xi\in\mathbb{R}$.
For $n \geqslant 2$, states of the form (\ref{DickeS2n}) satisfy Eqs.~(\ref{mngc1})--(\ref{mngc2}) and are thus all $1$-anticoherent.

\subsubsection{Dihedral groups}
Because $C_n$ is a subgroup of $D_n$, the Dicke coefficients of a configuration with symmetry group $D_n$ are of the form \eqref{DickeCn}, with the additional invariance under $\sigma_h\sigma_v$ given in Table~\ref{tabDickeconstraints}. These conditions give the canonical forms
\begin{multline}
\label{DickeDn1}
\mathbf{d}_{D_n} = (\mathbf{0}_{n_S},d_{n_S},\mathbf{0}_{n-1},d_{n_S+n},\mathbf{0}_{n-1},d_{n_S+2n},\ldots\\
\ldots,d_{n_S+2n},\mathbf{0}_{n-1},d_{n_S+n},\mathbf{0}_{n-1},d_{n_S},\mathbf{0}_{n_S})
\end{multline}
or
\begin{multline}
\label{DickeDn2}
\mathbf{d}_{D_n} = (\mathbf{0}_{n_S},d_{n_S},\mathbf{0}_{n-1},d_{n_S+n},\mathbf{0}_{n-1},d_{n_S+2n},\ldots\\
\ldots,-d_{n_S+2n},\mathbf{0}_{n-1},-d_{n_S+n},\mathbf{0}_{n-1},-d_{n_S},\mathbf{0}_{n_S}).
\end{multline}
For $n \geqslant 2$, states of the form (\ref{DickeDn2}) satisfy Eqs.~(\ref{mngc1})--(\ref{mngc2}) and are thus all $1$-anticoherent.

The $D_{nh}$ symmetry additionally imposes invariance under $\sigma_h$ given in Table~\ref{tabDickeconstraints}, which leads to the canonical forms
\begin{multline}
\label{DickeDnh}
\mathbf{d}_{D_{nh}} = (\mathbf{0}_{n_S},d_{n_S},\mathbf{0}_{n-1},d_{n_S+n},\mathbf{0}_{n-1},d_{n_S+2n},\ldots\\
\ldots, d_{n_S+2n},\mathbf{0}_{n-1},  d_{n_S+n},\mathbf{0}_{n-1},  d_{n_S},\mathbf{0}_{n_S}),  \\ d_k \in \mathbb{R}
\end{multline}
or
\begin{multline}
\label{DickeDnhm}
\mathbf{d}_{D_{nh}} = (\mathbf{0}_{n_S},d_{n_S},\mathbf{0}_{n-1},d_{n_S+n},\mathbf{0}_{n-1},d_{n_S+2n},\ldots\\
\ldots,- d_{n_S+2n},\mathbf{0}_{n-1}, - d_{n_S+n},\mathbf{0}_{n-1}, - d_{n_S},\mathbf{0}_{n_S}),\\  d_k \in \mathbb{R} .
\end{multline}
The $5$-anticoherent states that will be given in Eq.~(\ref{t4familyeq}) are examples of states displaying $D_{nh}$ symmetry.

The $D_{nd}$ symmetry imposes the form \eqref{DickeS2n} together with the invariance under $\sigma_d$ given in Table~\ref{tabDickeconstraints}, which implies the general condition that the $d_k$ be real and $d_{N-k}=(-1)^qd_k$ for indices $k=n_S+q n$, that is, the forms
\begin{multline}\label{DickeDnd}
\mathbf{d}_{D_{nd}} = (\mathbf{0}_{n_S},d_{n_S},\mathbf{0}_{n-1},d_{n_S+n},\mathbf{0}_{n-1},d_{n_S+2n},\ldots\\
\ldots, d_{n_S+2n},\mathbf{0}_{n-1},-d_{n_S+n},\mathbf{0}_{n-1},d_{n_S},\mathbf{0}_{n_S}), \\ d_k \in \mathbb{R}
\end{multline}
or
\begin{multline}\label{DickeDndm}
\mathbf{d}_{D_{nd}} = (\mathbf{0}_{n_S},d_{n_S},\mathbf{0}_{n-1},d_{n_S+n},\mathbf{0}_{n-1},d_{n_S+2n},\ldots\\
\ldots, -d_{n_S+2n},\mathbf{0}_{n-1},d_{n_S+n},\mathbf{0}_{n-1},-d_{n_S},\mathbf{0}_{n_S}), \\ d_k \in \mathbb{R}.
\end{multline}
The $7$-anticoherent state that will be presented in Fig.~\ref{n42t7} is an example of a state displaying $D_{7d}$ symmetry.

To summarize, $C_n$-symmetric states all have a diagonal $t$-qubit reduced density matrix for $t<n$, but are not necessarily $1$-anticoherent, whereas $C_{nh}$, $S_{2n}$ and $D_n$-symmetric states are always a least $1$-anticoherent. This includes states with Majorana points invariant under inversion (which transforms a point into its symmetric with respect to the center of the sphere), which corresponds to the symmetry group $S_2$.

\section{Anticoherent states in SLOCC classes}\label{sec:slocc}

Local unitaries are a special case of more general transformations used in the context of quantum information theory, namely stochastic local operation with classical communication (SLOCC). By definition, two states $\ket{\Psi}$ and $\ket{\Phi}$ are equivalent under SLOCC if there exists a local protocol with classical communication which transforms one state into the other with a finite probability of success. Mathematically, this corresponds to the requirement that these two states can be related by an invertible local operation (ILO) $A_1\otimes \ldots\otimes A_N$ with $A_k$ invertible operators~\cite{Dur}. This defines an equivalence relation which partitions the Hilbert space into different SLOCC classes. For symmetric states, the $A_k$ can be chosen equal~\cite{Math10}, so that two symmetric states $\ket{\Phi_S}$ and $ \ket{\Psi_S}$ belong to the same SLOCC class iff there exists an invertible operator $A$ acting on a single qubit such that $\ket{\psiS}=A^{\otimes N}\ket{\Phi_S}$. Each SLOCC class contains at most one state (up to LU) with a maximally mixed one-qubit reduced density matrix~\cite{Verstraete}. Thus, each SLOCC class contains at most one anticoherent state. The unicity (up to LU) of $1$-anticoherent states within their SLOCC class makes them natural representatives of SLOCC classes~\cite{Gour_13}. As a corollary, it follows that two anticoherent states which are not LU-equivalent necessarily belong to different SLOCC classes. Moreover the union of SLOCC classes containing a $1$-anticoherent state is dense in Hilbert space~\cite{Gour_13}.

The various SLOCC classes of symmetric states can be gathered into families denoted by $\mathcal{D}_{m_1, m_2,\ldots, m_d}$~\cite{Bas09}, which are defined by their diversity degree $d$ (the number of distinct Majorana points on the Bloch sphere), and by their degeneracy configuration $m_1, m_2,\ldots, m_d$ (the degeneracy of each Majorana points). For instance the family $\mathcal{D}_{1,1,\ldots,1}$ contains SLOCC classes with states such that all Majorana points are non degenerate and the family $\mathcal{D}_N$ contains a single SLOCC class (that of separable symmetric states). All families with a diversity degree $d\leqslant 3$ contain a single SLOCC class, since any set of three distinct points can be transformed into any other set of three distinct points by SLOCC transformation~\cite{Bas09,Aulbach10}.

\subsection{$C_n$ symmetry, $1$-anticoherence and SLOCC classes}\label{subsec:existence}

We now provide a way of finding, given a $C_n$-symmetric state, its SLOCC-equivalent $1$-anticoherent state (when it exists). 
Our result provides a clear link between symmetry and entanglement classes. We show that any SLOCC class belonging to a family $\mathcal{D}_{m_1, m_2,\ldots, m_d}$ with all $m_k< N/2$ and containing states with $C_n$ symmetry possesses a $1$-anticoherent state of $C_n$ symmetry. Moreover, any SLOCC class belonging to a family with at least one $m_k \geqslant N/2$ ($k=1,\ldots,d$) does not contain any anticoherent state, except for the family $\mathcal{D}_{N/2,N/2}$, for which $\ket{D_{N}^{(N/2)}}$ is $1$-anticoherent (but not $2$-anticoherent)~\cite{DTJ14}. This statement has the following corollary : if a SLOCC class contains a $1$-anticoherent state which does not display $C_n$ symmetry, then the class does not contain any state displaying $C_n$ symmetry (as anticoherent states are unique up to LU within their SLOCC class). Note that the possibility of a connection between symmetry and types of entanglement has been pointed out in~\cite{Mar11}.

A state displaying $C_n$ symmetry can always be brought by LU to the canonical form (\ref{DickeCn}). This state can be converted via a diagonal ILO $A^{\otimes N}$ with $A=\mathrm{diag}(y,1)$ to a SLOCC-equivalent state
\begin{multline}
\label{DickeCnNG}
\tilde{\mathbf{d}}_{C_n} =\mathcal{N}(\mathbf{0}_{n_S},d_{n_S},\mathbf{0}_{n-1},d_{n_S+n}\,y^{\frac{n}{2}},\\
\mathbf{0}_{n-1},d_{n_S+2n}\,y^{n},\ldots,\mathbf{0}_{n_N})
\end{multline}
with the same $C_n$ symmetry (throughout this work, $\mathcal{N}$ will denote a normalization constant). The state (\ref{DickeCnNG}) can at least be made $1$-anticoherent. It suffices to take $y$ equal to the only strictly positive root of the polynomial
\begin{equation}
\label{poly}
\sum_{k=0}^{N}(N-2k)|d_{k}|^2\,y^{k}.
\end{equation}
The existence and unicity of the root follows from Descartes' rule of signs : if the degeneracy of Majorana points is smaller than $N/2$, we are ensured that at least one $d_k$ and one $d_{k+N/2}$ for $k=0,\ldots, N/2$ are non-zero, so that there is exactly one change of sign in the coefficients of the polynomial~(\ref{poly}). 

Thus, from any $C_n$-symmetric state belonging to a family $\mathcal{D}_{m_1, m_2,\ldots, m_d}$ with all $m_k< N/2$, our procedure allows us to construct the SLOCC-equivalent $C_n$-symmetric $1$-anticoherent state. This is particularly useful since $1$-anticoherence is a requirement for $t$-anticoherence. Moreover, it provides a practical way of determining whether two states displaying cyclic symmetries in their Majorana points (with degeneracies $m_k< N/2$) belong to the same SLOCC class. It suffices to determine the point arrangements of their SLOCC-equivalent anticoherent states (\ref{DickeCnNG}), and check whether they are identical up to rigid rotation of the sphere. 

\subsection{Illustrations}

\subsubsection{Case $N\leqslant 4$}

All anticoherent states of up to $N=4$ qubits have been identified in~\cite{DTJ14}. Remarkably, they all display $C_n$ symmetry, and coincide with anticoherent states found using the method presented in the preceding subsection. For $2$ and $3$ qubits, the only $1$-anticoherent states (up to LU) are the $C_2$-symmetric Bell state $\mathbf{d}_{C_2} = \tfrac{1}{\sqrt{2}}(1,0,1)$ and the $C_3$-symmetric GHZ state $\mathbf{d}_{C_3} = \tfrac{1}{\sqrt{2}}(1,0,0,1)$. The case of $4$ qubits is more interesting as there is an infinite number of SLOCC-inequivalent $1$-anticoherent states of the form $\mathbf{d}_{C_2} = \mathcal{N}(1,0,\tau,0,1)$ with $\tau \in \mathbb{C}$~\cite{DTJ14}. Their Majorana representation displays a dihedral $D_2$ symmetry. Our method allows us to recover these states starting from the general form (\ref{DickeCn}) of a $4$-qubit $C_2$-symmetric state, $\mathbf{d}_{C_2} = \mathcal{N}(1,0,\mu,0,\nu)$ with $\nu>0$~\cite{footnote2}, and then using the SLOCC-equivalent state~(\ref{DickeCnNG}) with $y=1/\sqrt{\nu}$ the only positive root of the polynomial (\ref{poly}).

\subsubsection{Case $N=5$}\label{subsec:n5}

\begin{center}
\begin{figure*}
\includegraphics[width=0.95\textwidth,clip=true]{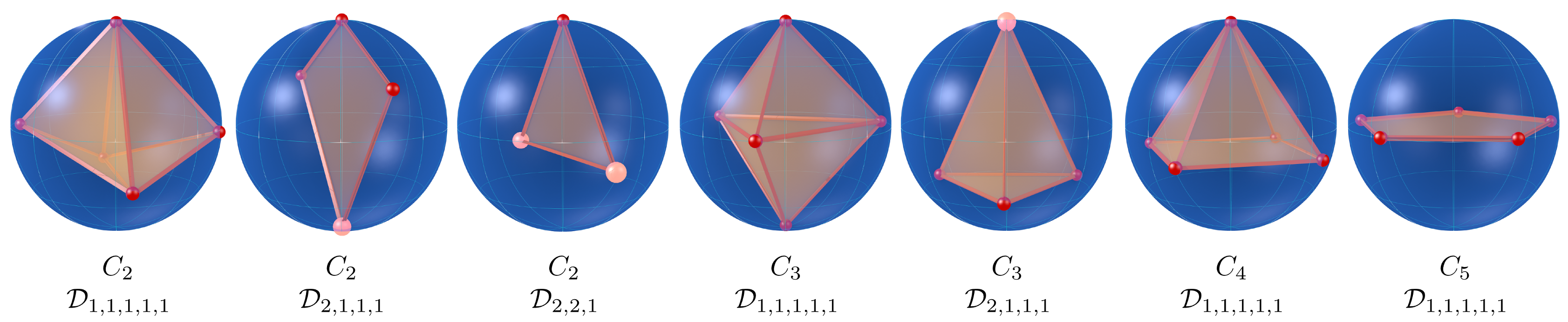}
\caption{(Color online) Majorana representation of 5-qubit 1-anticoherent states displaying $C_n$ symmetry. The small/large points on the spheres correspond to non-degenerate/twice-degenerate Majorana points. The corresponding states are those given in the main text (see Sec.~\ref{subsec:n5}).\label{img_N5}}
\end{figure*}
\end{center}

We consider here exhaustively the case of $5$-qubit states with any possible $C_n$ symmetry. As discussed in Sec.~\ref{subsec:existence}, classes belonging to families characterized by points with multiplicity $m\geqslant N/2$ do not contain any anticoherent states. On the other hand, all other classes belonging to the three families $\mathcal{D}_{1,1,1,1,1}$, $\mathcal{D}_{2,1,1,1}$, and $\mathcal{D}_{2,2,1}$ and containing $C_n$-symmetric states have exactly one (up to LU) anticoherent state. We now consider all SLOCC classes containing a $C_n$-symmetric state for some order $n$, and explicitly construct the related $1$-anticoherent state. It can be checked that none of these states is $2$-anticoherent.

\paragraph{$C_2$ symmetry.} 
If the $C_2$-symmetric state belongs to the family $\mathcal{D}_{1,1,1,1,1}$, all points are non-degenerate, which implies in particular that exactly one point must lie at a pole in order to fulfill the symmetry requirement. If we choose this point to lie at the North pole, Eq.~(\ref{DickeCn}) reduces to
\begin{equation}
\label{C2N5}
\mathbf{d}_{C_2} = \mathcal{N}(1,0,\mu,0,\nu,0),
\end{equation}
where $\mu>0$ and $\nu\in\mathbb{C}$~\cite{footnote2}. From Eq.~(\ref{DickeCnNG}) the $1$-anticoherent state which is  SLOCC-equivalent to (\ref{C2N5}) is of the form
\begin{equation}
\mathbf{d}_{C_2} = \mathcal{N}(1,0,\mu\,y,0,\nu\,y^2,0)
\end{equation}
with
\begin{equation}\label{eq:c11111n5ng}
y=\frac{1}{\sqrt{6}|\nu|}\sqrt{\mu^2+\sqrt{\mu^4+60|\nu|^2}}
\end{equation}
the only positive root of the polynomial (\ref{poly}). The Majorana representation of this state is shown in Fig.~\ref{img_N5} for $\mu = 2$ and $\nu = 10\, i$.

The most general state with symmetry $C_2$ in the family $\mathcal{D}_{2,1,1,1}$ takes (up to LU) the form $\mathbf{d}=\mathcal{N}(0,0,1,0,\mu,0)$ with $\mu\geqslant 0$, as the degenerate point is necessarily at a pole, which can be chosen as the South pole. As follows from Eq.~(\ref{DickeCnNG}) with $y=1/(\sqrt{3}\mu)$ the positive root of Eq.~(\ref{poly}), it can be brought to the $1$-anticoherent state $\mathbf{d}_{C_2}=(0,0,\sqrt{3},0,1,0)/2$, whose Majorana representation is shown in Fig.~\ref{img_N5}. Thus states with $C_2$ symmetry in the family $\mathcal{D}_{2,1,1,1}$ all belong to the same SLOCC class. 

The family $\mathcal{D}_{2,2,1}$ contains states which have five Majorana points, two of which are doubly-degenerate, leaving only three distinct points. Therefore, there is only one SLOCC class in the family $\mathcal{D}_{2,2,1}$ (see Sec.~\ref{sec:slocc}), which contains (up to LU) a $1$-anticoherent state displaying $C_2$ symmetry, namely $\mathbf{d}_{C_2}=\mathcal{N}(1,0,\mu,0,\nu,0)$ with $\mu=\sqrt{\tfrac{2}{15}(1+2\sqrt{19})}$ and $\nu=\tfrac{1}{15}(\sqrt{5}+2\sqrt{95})$. Indeed, it is easy to check that this state is of the form (\ref{DickeCn}) and verifies Eqs.~(\ref{mngc1}) and (\ref{mngc2}) (see Fig.~\ref{img_N5}). 

\paragraph{$C_3$ symmetry.} For the family $\mathcal{D}_{2,2,1}$, there is no way to satisfy the $C_3$ symmetry. Any $C_3$-symmetric state belonging to the family $\mathcal{D}_{1,1,1,1,1}$ has to be of the form $\mathbf{d}_{C_3} = \mathcal{N}(0,1,0,0,\mu,0)$, with one point at the North pole and one point at the South pole. Using Eq.~(\ref{DickeCnNG}), the SLOCC-equivalent $1$-anticoherent state is given by $\mathbf{d}_{C_3} = \tfrac{1}{\sqrt{2}}(0,1,0,0,1,0)$, actually displaying $D_{3h}$ symmetry by virtue of Eq.~(\ref{DickeDnh}) (see Fig.~\ref{img_N5}).

Similarly, one can show that $C_3$-symmetric states belonging to the family $\mathcal{D}_{2,1,1,1}$ are SLOCC-equivalent to the $1$-anticoherent state
$\mathbf{d}_{C_3} = \tfrac{1}{\sqrt{6}}(1,0,0,\sqrt{5},0,0)$ (see Fig.~\ref{img_N5}).

\paragraph{$C_4$ symmetry.} The only family with states having $C_4$ symmetry is $\mathcal{D}_{1,1,1,1,1}$ and their most general canonical form reads $ \mathbf{d}_{C_4} = \mathcal{N}(1,0,0,0,\mu,0)$, which can be brought by SLOCC to the $1$-anticoherent state $\mathbf{d}_{C_4} =\tfrac{1}{\sqrt{8}}(\sqrt{3},0,0,0,\sqrt{5},0)$  (see Fig.~\ref{img_N5}). 

\paragraph{$C_5$ symmetry.} The only family with states having $C_5$ symmetry is $\mathcal{D}_{1,1,1,1,1}$ and their most general canonical form reads $ \mathbf{d}_{C_5} = \mathcal{N}(1,0,0,0,0,\mu)$, which can be brought by SLOCC to the $1$-anticoherent state $\mathbf{d}_{C_5} = \tfrac{1}{\sqrt{2}}(1,0,0,0,0,1)$, actually displaying $D_{5h}$ symmetry as follows from Eq.~(\ref{DickeDnh}) (see Fig.~\ref{img_N5}).

\subsubsection{Exemple for $N=6$}

The case of $6$ qubits can be treated in the same way as the $5$-qubit case. As for $N=5$, families with four-fold or higher degeneracy of Majorana points do not contain any anticoherent state. Family $\mathcal{D}_{3,3}$ contains a single SLOCC class, in which $\ket{D_6^{(3)}}$ is the only (up to LU) $1$-anticoherent state. All other classes containing a $C_n$-symmetric state, belonging to the four families $\mathcal{D}_{1,1,1,1,1,1}$, $\mathcal{D}_{2,1,1,1,1}$, and $\mathcal{D}_{2,2,1,1}$, $\mathcal{D}_{2,2,2}$, each contain a unique (up to LU) $1$-anticoherent state. Moreover, for $N=6$, it is possible to find higher order anticoherent states, as we now show on an example.

\paragraph{$C_3$ symmetry.} We consider a state of the form $\mathbf{d}_{C_3}=\mathcal{N}(1,0,0,\mu,0,0,\nu)$ which is SLOCC-equivalent to the $1$-anticoherent state $\mathbf{d}_{C_3}=\mathcal{N}(1,0,0,\mu \,y^{3/2},0,0,\nu \,y^{3})$ with $y=1/|\nu |^{1/3}$. For the state to be $2$-anticoherent, it must additionally verify Eq.~(\ref{t2c1}), i.e.\ $\mu=\sqrt{\tfrac{5}{2}|\nu|}$, which leads to
\begin{equation}\label{C3AC}
\mathbf{d}_{C_3}=\frac{1}{3}(\sqrt{2},0,0,\sqrt{5},0,0,\sqrt{2}\,e^{i \varphi_\nu}).
\end{equation}
The only parameter left is $\varphi_\nu$, resulting in a one-parameter family of SLOCC-inequivalent $2$-anticoherent states (the SLOCC-inequivalence is a consequence of the LU-inequivalence which can be checked from the relative positions of the Majorana points). Among these states, the one with $\varphi_\nu=\pi$ is the only 3-anticoherent state (it verifies the additional off-diagonal conditions (\ref{nondiagatdk}) for $t=3$). It corresponds to an octahedral arrangement of Majorana points.

\paragraph{$C_4$ symmetry.} 

A $C_4$-symmetric state of $6$ qubits can be constructed by putting one Majorana point at the North pole and another at the South pole. The Dicke coefficients are then of the form $\mathbf{d}_{C_4}=\mathcal{N}(0,1,0,0,0,\mu,0)$ with $\mu>0$. This state is SLOCC-equivalent to the $1$-anticoherent state $\mathbf{d}_{C_4}=\frac{1}{\sqrt{2}}(0,1,0,0,0,1,0)$ (obtained by using $y=1/\sqrt{\mu}$ the positive root of Eq.~(\ref{poly})). This state is $3$-anticoherent, as can be easily checked through Eqs.~(\ref{diagatdk})--(\ref{nondiagatdk}). It corresponds to an octahedral arrangement of Majorana points, and is thus LU-equivalent to state (\ref{C3AC}) with $\varphi_\nu=\pi$. Note that this state has both $C_3$ and $C_4$ symmetry, and coincides with both the most quantum spin-$3$ state~\cite{Gir10}, and the most entangled $6$-qubit symmetric state with respect to the geometric measure of entanglement~\cite{Martin10}.

\section{Infinite families of higher order anticoherent states with $C_n$ symmetry}\label{sec:famillies}

A natural question is to ask whether there exist $t$-anticoherent states for arbitrary number of qubits $N$ and order $t$. For larger numbers of qubits, it becomes difficult to find analytical solutions to Eqs.~(\ref{diagatdk})--(\ref{nondiagatdk}). It is however possible to construct infinite families of anticoherent states. An example was already given in Sec.~\ref{Majorep}. Another family of $1$-anticoherent states was proposed in~\cite{DTJ14}, and reads
\begin{equation}
\frac{1}{\sqrt{2N-2}}\left(\sqrt{N-2}\ket{D_N^{(0)}}+\sqrt{N}\ket{D_N^{(N-1)}}\right).
\end{equation}
A family of $2$-anticoherent states was proposed in~\cite{Zimba_EJTP_3} for even $N$ with $N\geqslant 6$,
\begin{equation}\label{statesZimba}
\sqrt{\tfrac{N+2}{6N}}\left(\ket{D_N^{(0)}}+2\sqrt{\tfrac{N-1}{N+2}}\ket{D_N^{(N/2)}}+\ket{D_N^{(N)}}\right).
\end{equation}
The states (\ref{statesZimba}) are, in fact, 3-anticoherent for $N\geqslant 8$, as can be checked from Eq.~(\ref{diagatdk}), given that Eq.~(\ref{diagat}) for $r\ne 0$ is automatically satisfied because of the $C_{N/2}$ symmetry. 

\begin{center}
\begin{figure*}
\includegraphics[width=1.0\textwidth,clip=true]{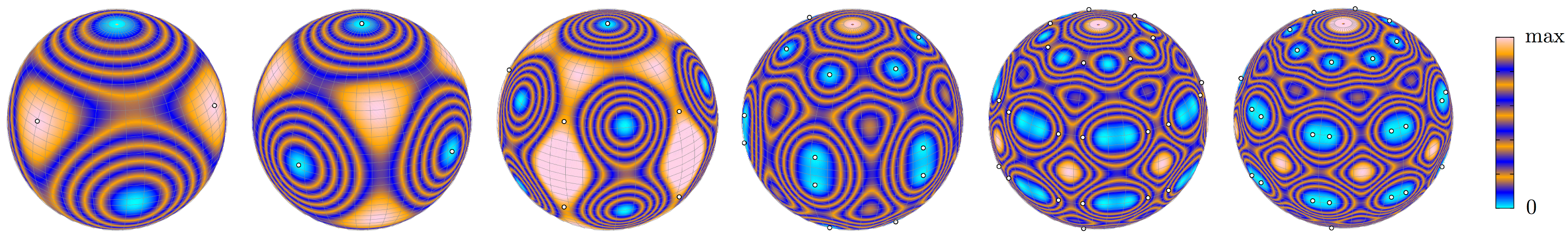}
\caption{(Color online) Husimi function of $C_n$-symmetric states corresponding to the smallest possible number $N_t$ of qubits for order of anticoherence $t=2,\ldots,7$ with $n>t$. From left to right: $2$-anticoherent state of $4$ qubits (tetrahedron, $C_3$ symmetry),  $3$-anticoherent state of $6$ qubits (octahedron, $C_4$ symmetry), $4$-anticoherent state of $12$ qubits ($C_5$ symmetry),  $5$-anticoherent state of $24$ qubits ($C_6$ symmetry),  $6$-anticoherent state of $42$ qubits ($C_7$ symmetry),  $7$-anticoherent state of $75$ qubits ($C_8$ symmetry).\label{img_gallery}}
\end{figure*}
\end{center}

We now present a systematic method which allows us, for a given order $t$ of anticoherence, to find $N$-qubit anticoherent states displaying $C_n$ symmetry with $n>t$ when they exist. As shown in Sec.~\ref{Canonical}, $C_n$ symmetry ensures that the off-diagonal equations~(\ref{nondiagatdk}) are satisfied for $t<n$. The remaining set of diagonal equations~(\ref{diagatdk}) can be reformulated as a linear system of equations in the variables $|d_{k}|^2$ with an additional positivity constraint on the solutions. Such a system can be solved analytically using linear programming. 

More specifically, for an $N$-qubit state with $C_n$ symmetry, the (normalized) vector of Dicke coefficients takes the form~(\ref{DickeCn}). Equation~(\ref{diagatdk}) becomes
\begin{equation}\label{seteq}
\sum_{k=0}^r \left(\frac{N}{2}-k \,n- n_S\right)^q |d_{n_S+k\, n}|^2 = A(q),
\end{equation}
where $r = (N-n_S-n_N)/n$ is the number of free Dicke coefficients, $n_S$ and $n_N$ are the number of points at the South and North poles respectively, and $A(q)$ is given by Eq.~(\ref{Aq}). By setting $x_k \equiv |d_{n_S + k \, n}|^2$ and $u_k = \frac{N}{2}-k \,n- n_S$, the set of equations  becomes
\begin{equation}\label{eq:systemn}
\begin{pmatrix}
1 & 1 & \cdots & 1 & 1 \\[3pt]
u_0 & u_1 & \cdots & u_{r-1} & u_{r} \\[3pt]
u_0^2 & u_1^2 &  & u_{r-1}^2 & u_{r}^2 \\[3pt]
\vdots & \vdots & \ddots  & \vdots & \vdots \\[3pt]
u_0^{t-1} & u_{1}^{t-1} & \cdots & u_{r-1}^{t-1} & u_{r}^{t-1} \\[3pt]
u_0^{t} & u_{1}^{t} & \cdots & u_{r-1}^{t} & u_{r}^{t}
\end{pmatrix}
\begin{pmatrix} 
x_0 \\[3pt]
x_1 \\[3pt]
x_2 \\[3pt]
\vdots \\[3pt]
x_{r-1} \\[3pt]
x_{r}
\end{pmatrix}
 = \begin{pmatrix} 
A(0) \\[3pt]
A(1) \\[3pt]
A(2) \\[3pt]
\vdots \\[3pt]
A(t-1)  \\[3pt]
A(t) \end{pmatrix}
\end{equation}
with $x_k\geqslant 0$. This system of $t+1$ equations and $r+1$ unknowns can be solved with linear programming, which guarantees to either find an analytic expression of a feasable solution or prove that there is no solution.

For fixed order of anticoherence  $t = 1, \ldots, 20$, we applied this method systematically for each $N = 1, \ldots, 500$ with all possible numbers of points $n_S$ and $n_N$ at the poles, and for all values of $n$ such that $t+1\leqslant n \leqslant N$ and $r = (N-n_S-n_N)/n$ be an integer. For a fixed order of anticoherence $t$, we found that a positive solution to Eq.~(\ref{eq:systemn}) exists only for $N$ larger than a certain value $N_t$. Examples of states with $N=N_t$ are represented in Fig.~\ref{img_gallery} for $t=2,\ldots,7$ through their Husimi function, showing that it becomes more and more uniform as $t$ increases.

\begin{figure}[h!]
\begin{centering}
\includegraphics[width=0.49\textwidth,clip=true]{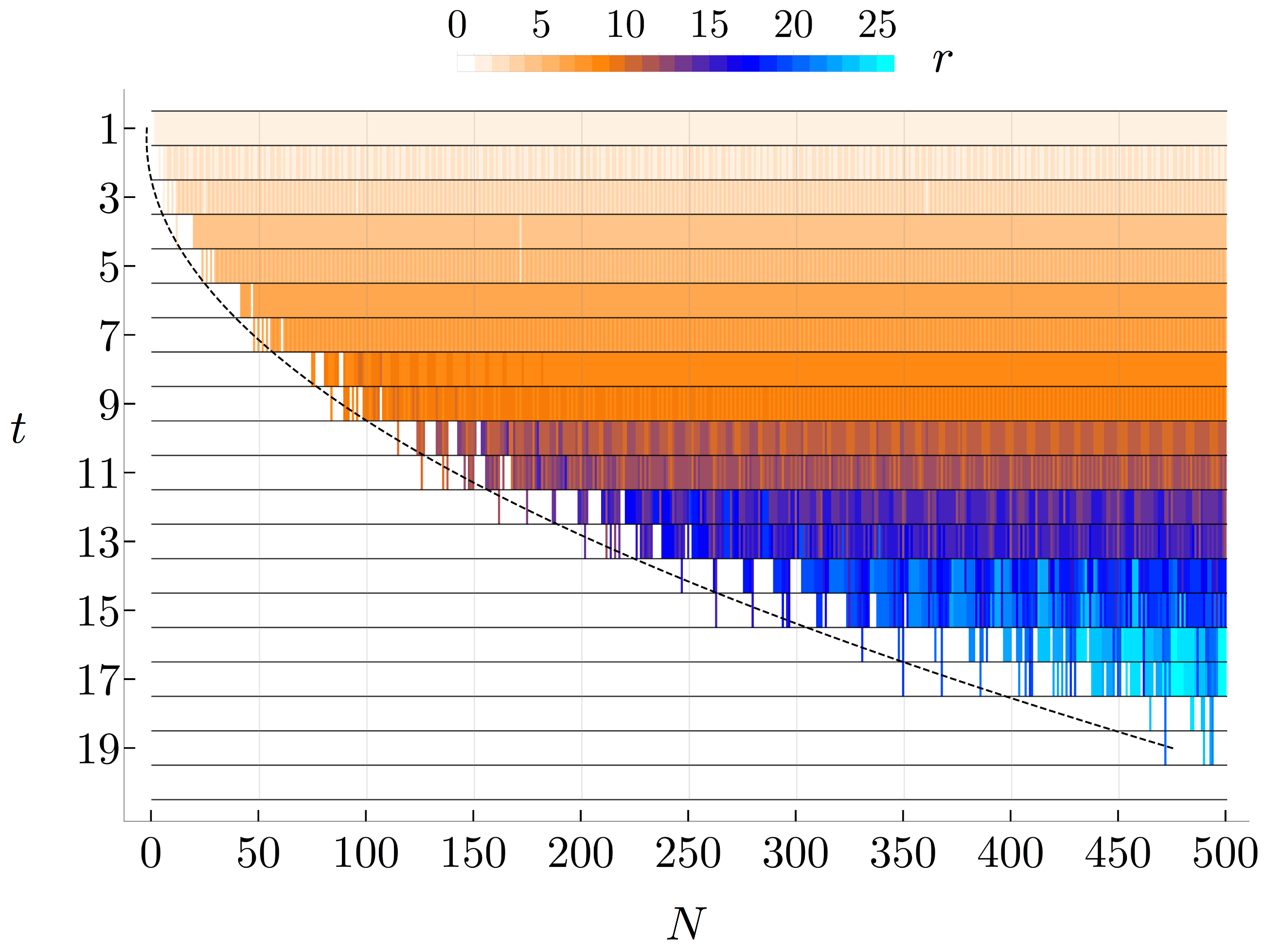}
\par\end{centering}
\caption{ (Color online) Each bar corresponds to a couple $(t,N)$ for which a $C_n$-symmetric $N$-qubit $t$-anticoherent state with $t<n$ exists. In particular, the minimum number of qubits $N_t$ for which such a state exists is given by the position of the first bar on each line.
This value $N_t$ is well fitted by $N_t=t (a t + b )$ with $a=1.52$ and $b=-3.93$ (dashed line). Color/greyness represents the number $r = (N-n_S-n_N)/n$ of free Dicke coefficients (all others being zero) on which the linear optimisation is realized. \label{rmin}}
\end{figure}

\begin{figure}
\begin{centering}
\includegraphics[width=5cm,clip=true]{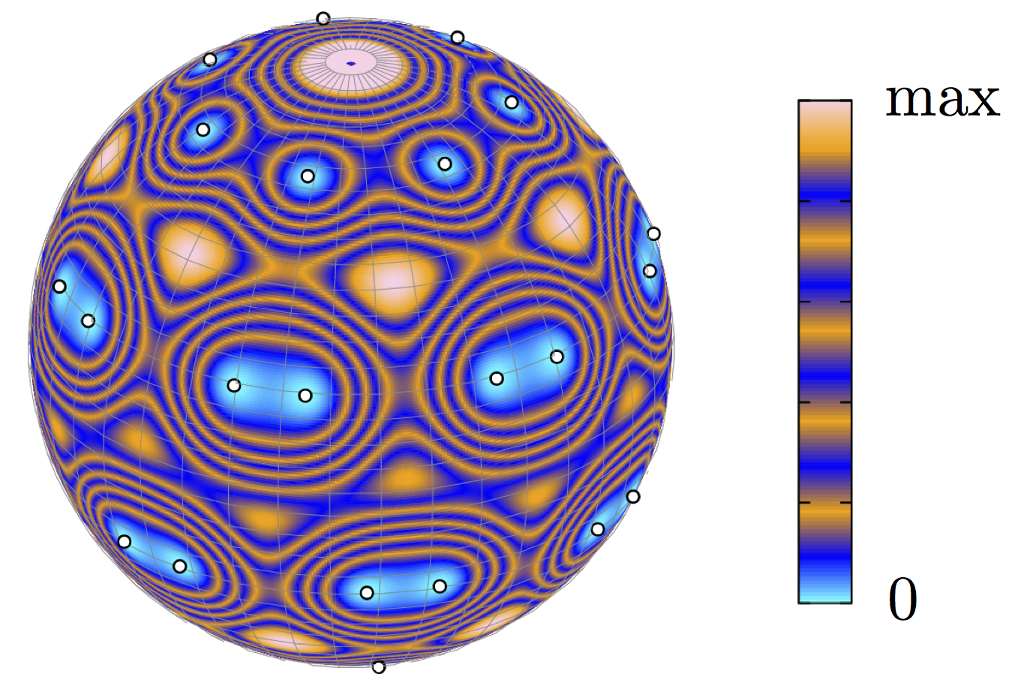}
\par\end{centering}
\caption{(Color online) Husimi function of the $42$-qubit $7$-anticoherent state $\ket{\psi_S}=(\sqrt{7062}\ket{D_{42}^{(0)}}+\sqrt{29315}\ket{D_{42}^{(7)}}+3\sqrt{451}\ket{D_{42}^{(14)}}+\sqrt{36777}\ket{D_{42}^{(21)}}-3\sqrt{451}\ket{D_{42}^{(28)}}+\sqrt{29315}\ket{D_{42}^{(35)}}-\sqrt{7062}\ket{D_{42}^{(42)}})/343$. This state displays $D_{7d}$ symmetry, hence $n=t=7$.
\label{n42t7}}
\end{figure}

Figure~\ref{rmin} displays all pairs $(t,N)$ for which $t$-anticoherent states of $N$ qubits with $C_n$ symmetry, $n>t$, are found to exist. As mentioned, no $t$-anticoherent states are found whenever $N$ is smaller than the threshold value $N_t$. As the figure indicates, the value of $N_t$ goes as $t^2$; this is to be expected since an $N$-qubit state has $N+1$ complex Dicke coefficients, which have to satisfy $(t+1)^2-1$ conditions for $t$-anticoherence. Above this minimal value $N_t$, $t$-anticoherent states can be found for almost all values of $N$. We stress that the bound $N_t$ is for states with $C_n$ symmetry such that $n>t$, ensuring that Eqs.~(\ref{diagat}) for $r\ne 0$ are trivially satisfied. These equations may however be satisfied in a non-trivial way when $n\leqslant t$. For instance, for $t=7$ we found a $42$-qubit state with $D_{7d}$ symmetry, displayed in Fig.~\ref{n42t7}. 

This approach enables us to identify infinite families of $t$-anticoherent states for higher values of $t$. For instance, a solution to (\ref{eq:systemn}) for $t=5$ and $N=4(m+1)$ with integer $m\geqslant 5$ is given by the $D_{m+1,h}$-symmetric states
\begin{equation}\label{t4familyeq}
\mathbf{d}_{D_{m+1,h}}=\mathcal{N}(d_0,\mathbf{0}_m,d_{N/4},\mathbf{0}_m,d_{N/2},\mathbf{0}_m,d_{3N/4},\mathbf{0}_m,d_{N})
\end{equation}
with
\begin{equation}
\begin{aligned}
& d_0=d_{N}=\sqrt{(2 + N) (4 + N) (7 N-4)},\\
& d_{N/4}=d_{3N/4}=4 \sqrt{2} \sqrt{(N-2) (N-1) (N+2)},\\
& d_{N/2}=2 \sqrt{3} \sqrt{(N-1) \left(N^2+16\right)}.\\
\end{aligned}
\end{equation}

We identified similar families for higher values of $t$.

\section{Anticoherence at all orders}\label{sec:allorder}
It is in fact possible, using the approach presented in this paper, to prove the existence of $t$-anticoherent states for arbitrary $t$. As we show here, one can construct $t$-anticoherent states with at most $t+1$ nonzero components. Let us consider a quantum state whose Dicke coefficients are zero for $k\neq k_i, 0\leq i\leq t$. We set $x_i=|d_{k_i}|^2$, so that the vector of Dicke coefficients takes the form $(0,\ldots,0,x_0,0,\ldots,0,x_1,0,\ldots,0,x_t,0,\ldots,0)$. The off-diagonal equations \eqref{nondiagatdk} are fulfilled as soon as there are $t$ or more zeroes between each $x_i$. Defining the functions $g_q(u)=(\frac12-u)^q$, we can rewrite Eq.~\eqref{diagatdk} as
\begin{equation}\label{diagatdk2}
\sum_{i=0}^t x_i\, g_q\!\left(\frac{k_i}{N}\right) =\frac{1}{N+1} \sum_{k=0}^{N}g_q\!\left(\frac{k}{N}\right).
\end{equation}
We want to find values of $k_i$ and $x_i>0$ such that \eqref{diagatdk2} holds, that is, sample points $k_i$ and positive weights $x_i$ such that the sum on the right-hand side, which runs from $k=0$ to $N$, can be evaluated by calculating the function $g_q$ at only $t+1$ points. As the right-hand side of Eq.~\eqref{diagatdk2} is a Riemann sum, which converges to the integral of $g_q$ over $[0,1]$ at large $N$, the problem almost appears like a quadrature problem.

The well-known Gauss-Legendre integration method~\cite{Hil86} states that the integral of an arbitrary function $g$ which is continuous over $[-1,1]$ can be approximated by evaluating $g$ at a finite number of points with a certain weight. The quadrature at order $t$ reads
\begin{equation}
\label{GaussLegendre}
\sum_{i=0}^{t}w_i\, g(u_i)\approx\int_{-1}^{1}g(u)du,
\end{equation}
where the $u_i$, $0\leq i\leq t$, are the $t+1$ roots of the Legendre polynomial $P_{t+1}(u)$, and weights are given by
\begin{equation}
\label{weightsGL}
w_i=\frac{2}{(t+1)P_{t+1}'(u_i)P_t(u_i)}.
\end{equation}
The Gauss-Legendre quadrature has the property that the approximation \eqref{GaussLegendre} becomes an exact equality when $g$ is a polynomial of degree $\leqslant 2t+1$. For functions defined over $[0,1]$ and for polynomials $g_q$ defined above, which are of degree less than $t$, the quadrature can be expressed, after a simple change of variables, as
\begin{equation}
\label{GaussLegendre01}
\sum_{i=0}^{t}\frac{w_i}{2}\, g_q\!\left(\frac{1+u_i}{2}\right)=\int_{0}^{1}\!g_q(u)du.
\end{equation}
As the weights \eqref{weightsGL} are such that $\sum_{i=0}^{t}w_i=2$, the weights $w_i/2$ in \eqref{GaussLegendre01} sum up to 1. The right-hand side of Eq.~\eqref{GaussLegendre01} is exactly the $N\to\infty$ limit of the right-hand side of Eq.~\eqref{diagatdk2}. Thus, choosing $k_i/N$ as close as possible to the roots $(1+u_i)/2$ should allow to solve \eqref{diagatdk2}.

To prove this, let us rewrite the system of equations \eqref{diagatdk2} in matrix form. In order to make the $N$-dependance clear we denote $X^{(N)}=(x_0,\ldots,x_t)$, $B^{(N)}=(A(0),\ldots,A(t))/N^q$ and define the matrix
\begin{equation}
M^{(N)}_{qi}=g_q\!\left(\frac{k_i}{N}\right),\qquad 0\leq q,i\leq t.
\end{equation}
Positions $k_i$ are chosen as
\begin{equation}
\label{defki}
k_i=\lfloor N (1+u_i)/2 \rfloor,\qquad 0\leq i\leq t,
\end{equation}
with $\lfloor . \rfloor$ the floor function. Equation \eqref{diagatdk2} reduces to $M^{(N)}X^{(N)}=B^{(N)}$. Similarly we can set $X^{(\infty)}=(w_0/2,\ldots, w_t/2)$, 
\begin{equation}
\label{defBinf}
B^{(\infty)}=\left(\int_{0}^{1}\!g_0(u)du,\ldots,\int_{0}^{1}\!g_t(u)du\right),
\end{equation}
and
\begin{equation}
M^{(\infty)}_{qi}=g_q\!\left(\frac{1+u_i}{2}\right)=\left(-\frac{u_i}{2}\right)^q,\quad 0\leq q,i\leq t,
\end{equation}
so that Eq.~\eqref{GaussLegendre01} for $q=0,\ldots,t$ can be reexpressed in matrix form as $M^{(\infty)}X^{(\infty)}=B^{(\infty)}$. The matrix $M^{(\infty)}$ is invertible since it is the Vandermonde matrix of the $-u_i/2$, which are all distinct, and thus $X^{(\infty)}=(M^{(\infty)})^{-1}B^{(\infty)}$. Since $k_i/N\to (1+u_i)/2$ for $N\to\infty$, we have $ M^{(N)}\to M^{(\infty)}$, and thus for sufficiently large $N$, the matrix $M^{(N)}$ is also invertible (note that the size $t+1$ of the matrix is fixed independently of $N$). Thus, for $k_i$ given by \eqref{defki} the system \eqref{diagatdk2} has a unique solution given by $X^{(N)}=(M^{(N)})^{-1}B^{(N)}$. In order to solve the diagonal equations (\ref{diagatdk}), it suffices to show that this solution is such that $x_i>0$. But for $N\to\infty$ we have $X^{(N)}=(M^{(N)})^{-1}B^{(N)}\to (M^{(\infty)})^{-1}B^{(\infty)}=X^{(\infty)}$, or equivalently, component by component, $x_i\to w_i/2$. Since the weights \eqref{weightsGL} of the Gauss-Legendre quadrature are all strictly positive, then for sufficiently large $N$ all $x_i$ will be positive. 

The off-diagonal equations \eqref{nondiagatdk} are fulfilled as soon as there are $t$ or more zeroes between each $x_i$. Since $t$, and thus the positions of the zeroes of $P_{t+1}(u)$, are fixed, then for sufficiently large $N$, indices defined by \eqref{defki} will be such that $\min_i|k_{i+1}-k_i|\geq t+1$. Thus a $t$-anticoherent state exists for any $t$.

Note that Gauss-Legendre quadrature with $t+1$ points is exact for any polynomial of degree less than or equal to $2t+1$. Thus the solution we construct is in fact $(2t+1)$-anticoherent. This means that we do not need to impose $t+1$ positions: only half of them would suffice. We can thus look for a solution with $D_{nh}$ symmetry, i.e.~such that the $k_i$ verify $k_i=k_{t-i}$ and $x_i=x_{t-i}$. Since the equations for odd $q$ are automatically fulfilled, there remains a set of (for $t$ even) $t/2+1$ equations and $t/2+1$ variables $x_0,\ldots,x_{t/2}$. The systems then corresponds to a quadrature to order $t/2$, which is exact for polynomials up to degree $t+1$. It is thus possible to obtain $D_{nh}$-symmetric $t$-anticoherent states for any $t$. 

The proof above only states the existence of some $N$ sufficiently large such that a $t$-anticoherent state exists. We were able to find solutions using $N=t(t+1)(t+2)/6$ up to $t=120$. More specifically, using linear programming, we looked for solutions of the form $(0,\ldots,0,x_0,0,\ldots,0,x_1,0,\ldots,0,x_t,0,\ldots,0)$ with $x_i=x_{t-i}$ and $k_i=k_{t-i}$, taking $k_i$ as in \eqref{defki} for $0\leq i\leq t/2$. We checked that for $t\geq 30$ the conditions $\min_i|k_{i+1}-k_i|\geq t+1$ (and thus the off-diagonal equations \eqref{nondiagatdk}) are always fulfilled; for even $t$ up to $t=120$, linear programming found an analytical solution satisfying all diagonal equations for $q=0,\ldots,t$.

\section{Conclusion}

In this paper, the problem of identifying $t$-anticoherent states of a spin-$j$ system (or equivalently $2j$-qubit permutation-symmetric states with maximally mixed $t$-qubit reductions in the symmetric subspace) has been tackled from a geometric point of view.  We have provided three different characterizations of $t$-anticoherence. A first one is given in terms of a finite number of expectation values of spin operators and is independent of any basis, a second one in terms of Dicke coefficients taking the form of a set of $t(t+2)$ real equations and a third one in terms of spherical multipole moments of the Husimi and Glauber-Sudarshan functions. 
Using the Majorana representation of a spin-$j$ state in terms of points on the sphere, we analyzed the consequence of a point-group symmetry of the Majorana points on the coefficients of the quantum state, for the seven infinite families of point group symmetries. Our results are summarized in Table~I. A consequence is that states with $C_{nh}$, $S_{2n}$ or $D_n$ symmetry are always at least $1$-anticoherent, in contrast to $C_n$-symmetric states which are not necessarily $1$-anticoherent. However, we were able to show that any SLOCC class containing $C_n$-symmetric states does contain a $C_n$-symmetric anticoherent state (provided the maximal degeneracy of its Majorana points is not too high). Because a SLOCC class contains at most one anticoherent state (up to LU), our results provide a clear link between symmetry and SLOCC classes. This approach allowed us to identify all anticoherent states with $C_n$ symmetry for $N=5$ qubits, and to construct families of $C_n$-symmetric states with higher order of anticoherence. We discussed a method based on linear programming allowing us to obtain $N$-qubit $C_n$-symmetric $t$-anticoherent states with $t<n$, when they exist. Finally we constructed explicitly $t$-anticoherent states with arbitrary order $t$, proving that such states exist at all order. Note that our results about anticoherence of spin states are of interest in the context of quantum information as the states we identified are maximally entanglement symmetric states with respect to the entanglement entropy.

\acknowledgments
FD would like to thank the F.R.S.-FNRS for financial support. FD is a FRIA grant holder of the Fonds de la Recherche Scientifique-FNRS. JM is grateful to the University of Li\`{e}ge for funding (project C-13/86).

\appendix
\section{Consequences of symmetries of the set of Majorana points on the Dicke coefficients}

\subsubsection{Reflection $\sigma_h$}
\label{subsubsh}
The reflection $\sigma_h$ transforms $z$ into $1/z^*$ and exchanges the North and South poles, so that a configuration of points invariant under $\sigma_h$ must be such that $n_N=n_S$ and
\begin{equation}
\label{productsh3}
\prod_{k=1}^{N-n_N-n_S}\left(z-z_k\right)=\prod_{k=1}^{N-2n_N}\left(z-\frac{1}{z_k^*}\right).
\end{equation}
We have
\begin{equation}
\label{productsh2}
\prod_{k=1}^{N-2n_N}\left(z-\frac{1}{z_k^*}\right)=\frac{(-z)^{N-2n_N}}{\prod_{k=1}^{N-2n_N}z_k^*}
\prod_{k=1}^{N-2n_N}\left(\frac{1}{z}-z_k^*\right).
\end{equation}
The coefficient of the lowest-order term in $z$ in Eqs.~\eqref{MajoranaPolynomialIdentity1}--\eqref{MajoranaPolynomialIdentity2} yields (for $n_S=n_N$) the identity
\begin{equation}
\prod_{k=1}^{N-2n_N}z_k=\frac{d_{n_N}}{d_{N-n_N}},
\end{equation}
so that Eq.~\eqref{productsh3} can be rewritten as
\begin{equation}
\frac{(-1)^N}{d_{n_N}^*} z^N P^*\left(\frac{1}{z}\right)=\frac{1}{d_{N-n_N}}P(z)
\end{equation}
(here $P^*$ is the polynomial whose coefficients are the complex conjugates of those of $P$). Identifying the coefficients of $z^k$ in the expansion yields that for all $k$ with $n_N\leq k\leq N-n_N$
\begin{equation}
d_{N-k}=\frac{d_{n_N}}{d_{N-n_N}^*}d_k^*.
\end{equation}
In particular, taking $k=n_N$ yields $|d_{N-n_N}|^2=|d_{n_N}|^2$. If we let $d_{N-n_N}/d_{n_N}^*=\exp(i\xi)$ then $d_{N-k}=e^{i\xi}d_k^*$. The condition of invariance under $\sigma_h$ is thus that $n_N=n_S$ and that there exists $\xi\in\mathbb{R}$ such that $d_{N-k}=e^{i\xi}d_k^* \;\forall\,  k$.

\subsubsection{Reflection $\sigma_v$}
The reflection $\sigma_v$ transforms $z$ into $z^*$. A configuration of points invariant under $\sigma_v$ must be such that
\begin{equation}
\prod_{k=1}^{N-n_N-n_S}\left(z-z_k\right)=\prod_{k=1}^{N-n_N-n_S}\left(z-z_k^*\right).
\end{equation}
Following the same lines as above, this yields the condition
\begin{equation}
\frac{1}{d_{N-n_N}^*} P^*(z)=\frac{1}{d_{N-n_N}}P(z),
\end{equation}
which by identification of the coefficients gives that there exists $\xi\in\mathbb{R}$ such that for all $k$ with $n_S\leq k\leq N-n_N$ one has $d_k^*=\exp(i\xi)d_k$. In particular this implies (iterating twice the condition) that $\xi=0$ or $\pi$, independently of $k$. One can further remove the overall sign of the $d_k$, leading to the conditions given in Table~\ref{tabDickeconstraints}.

\subsubsection{Reflection $\sigma_d$}
The reflection $\sigma_d$ transforms $z$ into $z^*\exp(-i\pi/n)$. Equation \eqref{polycn} becomes
\begin{equation}
\label{polysd}
P(z)=e^{-\frac{i\pi}{n}(N-n_N)}\frac{d_{N-n_N}}{d_{N-n_N}^*}P^*\left(ze^{\frac{i\pi}{n}}\right)
\end{equation}
so that 
\begin{equation}
\label{dkrn}
d_k=e^{-\frac{i\pi}{n}(N-n_N-k)}\frac{d_{N-n_N}}{d_{N-n_N}^*}d_k^*
\end{equation}
for all $k$ with $n_S\leq k\leq N-n_N$. If we let $e^{-\frac{i\pi}{n}(N-n_N)}d_{N-n_N}/d_{N-n_N}^*=e^{i\xi}$ then Eq.~\eqref{dkrn} yields
\begin{equation}
d_k=e^{i\xi}e^{\frac{ik\pi}{n}}d_k^*.
\end{equation}
This imposes for the phase $\theta_k$ of $d_k$ that $\theta_k=\xi/2+k\pi/n+q_k\pi$ for some integer $q_k$. Removing the overall constant phase, the $d_k$ can be taken as
\begin{equation}
d_k=\pm |d_k|e^{\frac{ik\pi}{2n}}.
\end{equation}

\subsubsection{Rotation $\sigma_h\sigma_v$}
This is a rotation of $\pi$ around the $x$-axis. It transforms $z$ into $1/z$. Invariance under $\sigma_h\sigma_v$ implies that $n_N=n_S$ and 
\begin{equation}
\label{productshsv}
\prod_{k=1}^{N-n_N-n_S}\left(z-z_k\right)=\prod_{k=1}^{N-2n_N}\left(z-\frac{1}{z_k}\right).
\end{equation}
Following the same steps as for $\sigma_h$, we get the condition
\begin{equation}
d_{N-k}=\frac{d_{n_N}}{d_{N-n_N}}d_k
\end{equation}
for all $k$ with $n_N\leq k\leq N-n_N$. Taking this expression at $k=n_N$ we have $d_{n_N}^2=d_{N-n_N}^2$, so that $d_{n_N}/d_{N-n_N}=\pm 1$, leading to the conditions given in Table~\ref{tabDickeconstraints}.

\subsubsection{Transformation $s_n=r_{2n}\sigma_h$}
Such an operation transforms $z$ into $\frac{1}{z^*}\exp{(-i\pi/n)}$. A configuration of points invariant under $s_n$ must be such that $n_N=n_S$ and
\begin{equation}
\label{productsn}
\prod_{k=1}^{N-n_N-n_S}\left(z-z_k\right)=\prod_{k=1}^{N-n_N-n_S}\left(z-\frac{1}{z_k^*}e^{-i\pi/n}\right).
\end{equation}
The identity that corresponds to this case can be derived as above and gives, for all $k$, $n_N\leq k\leq N-n_N$,
\begin{equation}
\label{rulesn}
d_{N-k}=\frac{d_{n_N}}{d_{N-n_N}^*}d_k^*e^{i\pi (N-k-n_N)/n}.
\end{equation}
Applying this condition to $k=n_N$ we get that $N-2n_N=0$ mod $n$. Iterating twice condition \eqref{rulesn} we get that $d_k=0$ unless $k=n_N$ mod $n$, which is the condition of invariance under $r_n$. This is to be expected as one has precisely $s_n^2=r_n$. If $k=n_N+q n$ for some integer $n$, and if  we let $d_{N-n_N}/d_{n_N}^*=\exp(i\xi)$, then \eqref{rulesn} becomes $d_{N-k}=(-1)^q \exp(i\xi)d_k^*$.

%%%%%%%%%%%%%%%%%%%%%%%%%%%%%%%%%%%%%%%%%%%%%%%%%%%%%%%%%%%%%%%%%%%%%%%%%%%%%%%
%******************************************************************************

%

\begin{thebibliography}{99}
\bibitem{footnote2} Two arbitrary non-zero Dicke coefficients of a state can always be chosen to be real and positive by applying a diagonal local unitary $U^{\otimes N}$ with $U=\mathrm{diag}(e^{i\alpha},1)$ to remove the relative phase between these two coefficients, and subsequently removing a global phase from the state.
\end{thebibliography}
\end{document}